\documentclass[a4paper,fleqn]{cas-sc}

\usepackage{natbib}
\setcitestyle{authoryear,open={(},close={)}} 

\usepackage{booktabs}
\usepackage{multirow}

\usepackage{lineno}

\def\tsc#1{\csdef{#1}{\textsc{\lowercase{#1}}\xspace}}
\tsc{WGM}
\tsc{QE}
\tsc{EP}
\tsc{PMS}
\tsc{BEC}
\tsc{DE}

\begin{document}
\let\WriteBookmarks\relax
\def\floatpagepagefraction{1}
\def\textpagefraction{.001}

\shorttitle{Elucidation of Unique Developmental Mechanism of Storm Surge along Northern Coast of Kyushu Island, Japan}

\shortauthors{Ozaki et~al.}

\title [mode = title]{Elucidation of Unique Developmental Mechanism of Storm Surge along Northern Coast of Kyushu Island, Japan}                      



%
\author[1]{Shinichiro Ozaki}

\cormark[1]


\ead{ozaki.shinichiro.247@s.kyushu-u.ac.jp}



\affiliation[1]{organization={Graduate School of Engineering, Kyushu University},
    addressline={Room W2-1003, 744 Motooka, Nishi-ku, Fukuoka City, Fukuoka Prefecture, Japan}, 
    postcode={819-0395}, 
    }

\author[2]{Yoshihiko Ide}
\ead{niimi-ms@n-koei.jp}
\affiliation[2]{
    addressline={Disaster Risk Reduction Research Center, Graduate School of Engineering, Kyushu University, Fukuoka, Japan}}

\author[3]{Masaki Niimi}
\ead{niimi-ms@n-koei.jp}
\affiliation[3]{
    addressline={Port \& Coast Department, Port \& Airport Division, NIPPON KOEI, Tokyo, Japan}}

\author[2]{Masaru Yamashiro}
\ead{rishi@stmdocs.in}

\author[4]{Mitsuyoshi Kodama}
\ead{m.kodama@et.kyushu-u.ac.jp}
\affiliation[4]{
    addressline={Technical Division, School of Engineering, Kyushu University, Fukuoka, Japan}}

\cortext[cor1]{Corresponding author}



\begin{abstract}
Along the northern coast of Kyushu Island, significant storm surges were unlikely to occur because the strong wind does not blow directly to the coast when typhoons passes. However, during Typhoon Maysak, various areas along the coast experienced flooding due to the storm surges. Additionally, inundation occurred when the typhoon was more than 600 km away from the coast. In this study, we classified the past typhoons into northeastward-moving, northward-moving and direct-passing overhead types and analyzed the storm surge using observational data and numerical simulations. Regarding northeastward-moving types, Hakata Bay located in the coast experienced two surge peaks. The first peak was induced by the inverted barometer effect and the stagnation of seawater in the Tsushima Strait. The second peak occurred because of the 10-hour oscillation and Ekman transport in the Tsushima Strait. For northward-moving types, Ekman transport was further intensified, resulting in a high storm surge that lasted for more than 10 hours. Regarding directly passing overhead types, one or two peaks occurred in a short period during the closest approach. The first peak was caused by the inverted barometer effect and Ekman transport, whereas the second peak was caused by the 2-hour harbor oscillation in Hakata Bay.
\end{abstract}


\begin{highlights}
\item Typhoons are categorized into different movement types and analyzed storm surges using observational data and simulations.
\item Northeastward-moving typhoons induced two surge peaks, driven by the inverted barometer effect and Tsushima Strait oscillation.
\item Northward-moving typhoons intensified Ekman transport, resulting in prolonged, high storm surges lasting over 10 hours.
\item Directly passing overhead typhoons caused one or two peaks due to harbor oscillations and inverted barometer effect effects.
\end{highlights}


\maketitle
\section{Introduction}\label{Introduction}

\begin{figure}
    \centering
    \includegraphics[width=\linewidth]{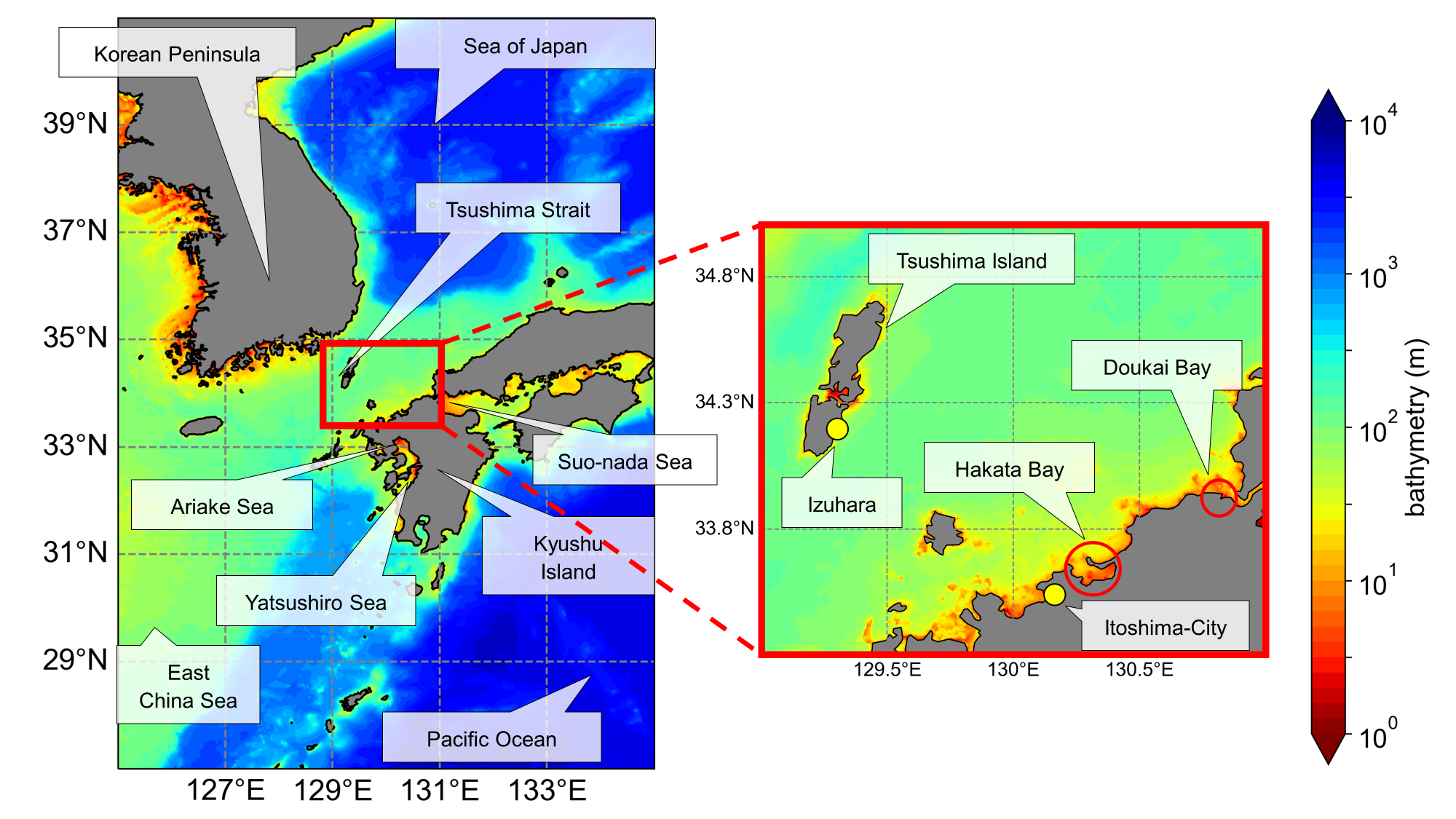}
    \caption{Bathymetry and local names of the northern coast of Kyushu Island.}
    \label{fig:target_sea_areas}
\end{figure}

Japan is prone to frequent typhoons and has experienced extensive storm surge damage. In particular, certain locations are known for their southward-facing bays, namely the three major bays along the Pacific coast—Tokyo Bay, Ise Bay, and Osaka Bay, referred to as the Three Great Bays—as well as the coastal areas of Kyushu Island, including the Suo-nada Sea, Ariake Sea, and Yatsushiro Sea (Figure~\ref{fig:target_sea_areas}). Indeed, their position leads to significant wind setup effects. Huge storm surges have been observed in the inner parts of these bays, and each of the said bays has a historical record of experiencing significant damage in the past (\citet{Yamamoto2021} and \citet{Mori2019}). As these coastal areas are densely populated, storm surge flooding can cause severe human and economic damage. Consequently, many studies have focused on these bays, with detailed investigations into the storm surge development characteristics of each bay (\cite{Nakajo2015}, \cite{Nakajo2017} and \cite{Ide2020}).
\begin{figure}
    \centering
    \includegraphics[width=0.5\linewidth]{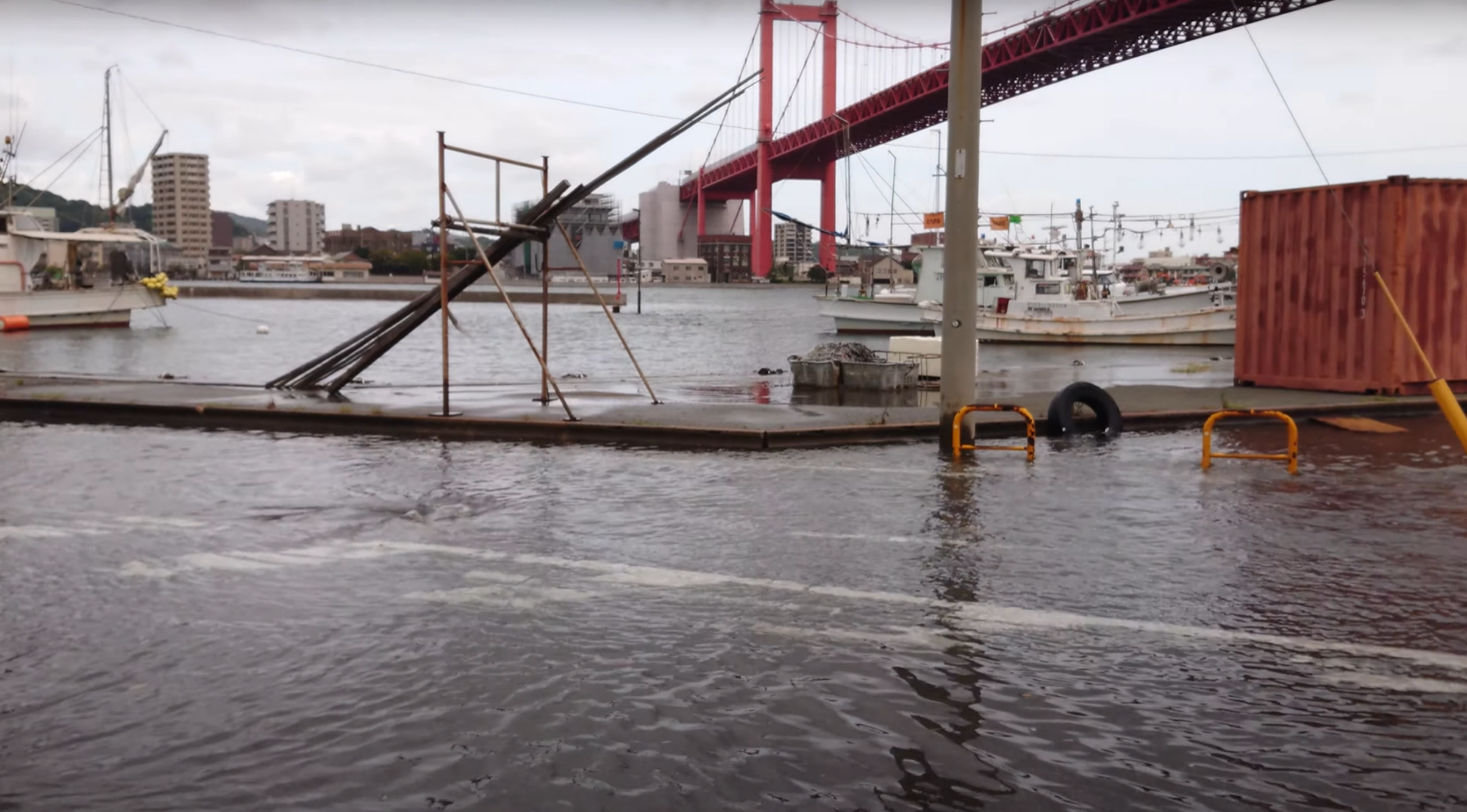}
    \caption{Inundation at the Dokai Bay during the impact of T2009. (\url{https://www.youtube.com/watch?v=G4IP9eHaw8Y})}
    \label{fig:Inundation_Dokai_Bay}
\end{figure}

Along the Sea of Japan and the Tsushima Strait, the bays face northward, resulting in a less prominent wind setup with regard to typhoons, and making it less likely that significant storm surges will occur. Historically, there have been few storm surges in these regions compared with the Three Great Bays and other bays on Kyushu Island. However, in recent years, the intensity of typhoons such as Maysak (T2009), Haishen (T2010), and Nanmadol (T2214) has increased. With this strengthening of typhoon intensity expected to continue owing to global warming (\cite{Yoshida2017} and \cite{Mori2022}), there is an increasing risk of storm surge disasters, necessitating consideration of the potential for inundation, even in regions that have not been well studied.

Indeed, during the impact of T2009, flooding was observed at multiple locations along the northern coast of Kyushu Island, such as Dokai Bay and Itoshima City (Figure~\ref{fig:Inundation_Dokai_Bay}). These points were previously unnoticed vulnerabilities to storm surges, where damage had not been confirmed in the past (Figure~\ref{fig:target_sea_areas}). Furthermore, inundation occurred when the typhoon had already moved toward the east coast of the Korean Peninsula and was located more than 600~km from the north coast of Kyushu Island. As a result, atmospheric pressure returned to normal levels, winds were calm, and the inverted barometer and wind setup effects were less likely to occur. The reason inundation occurred after the passage of the typhoon remains unclear.

In this study, we focused on the unique developmental processes of storm surges along the northern coast of Kyushu Island, adjacent to the largest city on the Sea of Japan side—an area that has not been extensively studied in the past. Previous research on storm surges along the northern coast of Kyushu includes studies by \cite{Hong1992}, \cite{Konishi2010}, \cite{Yamashiro2016}, \cite{Niimi2022}, and \cite{Ide2023}.
\cite{Hong1992} investigated the time-varying flow and storm surge around the Tsushima Strait during typhoon passages, using observed values and numerical simulations in Hakata, Izuhara, and Busan (Figure~\ref{fig:target_sea_areas}). 
\cite{Konishi2010} focused on Seibolt Typhoon in 1828, which is considered the most powerful typhoon to have occurred in the past 300 years. Through historical documents, observational data, and numerical simulations, this study investigated the realities and causes of disasters and typhoon intensity, revealing that the maximum storm surge heights in the Ariake Sea, Suo-nada, and Hakata Bay exceeded 4.5~m, 3.5~m, and 3~m, respectively. Furthermore, \cite{Yamashiro2016} conducted a detailed examination of the characteristics of storm surges in Hakata Bay using numerical simulations with high spatial resolution. The aforementioned study revealed the variation characteristics of storm surges based on spatial and temporal fluctuations in wind direction, wind speed, and atmospheric pressure.
\cite{Niimi2022} focused on the flooding case in Dokai Bay during T2009, examining its mechanism. The study concluded that the flooding was caused by the overlap of storm surges due to Ekman transport during high tide.
In coastal areas where strong winds do not blow directly, Ekman transport along the coastline due to winds parallel to the shore is known to be the primary cause of storm surges, as confirmed in various regions worldwide, including Japan (\cite{Shen2009}, \cite{Hashimoto2007}, \cite{Kim2010} and \cite{Dieng2021}).
\cite{Ide2023} investigated the characteristics of anomalies, including normal conditions, along the Tsushima Strait coast throughout the year, using observational data and numerical simulations. The said study revealed that anomalies along the strait were strongly influenced by Ekman transport, highlighting the distinct spatial features between wind setup effects and Ekman transport. 

Research on storm surges driven by Ekman transport has primarily focused on open-ocean areas. However, the northern coast of Kyushu Island, the subject area of the present study, faces the Tsushima Strait, with the Korean Peninsula located approximately 200~km away as shown in Figure \ref{fig:target_sea_areas}. The strait connects the East China Sea and Sea of Japan at its ends, creating an area with extreme variations in water depth (Figure~\ref{fig:target_sea_areas}).
Storm surge development mechanisms in such complex regions are expected to exhibit characteristics distinct from those observed in the open ocean areas targeted by previous studies. Although \cite{Hong1992} and \cite{Niimi2022} investigated the storm surge characteristics in areas facing the Tsushima Strait, their studies were limited to specific typhoon events or periods.
As such, we explored multiple typhoons, categorized them into various trajectory patterns, and examined the storm surge development characteristics associated with each trajectory pattern. Additionally, this study provides explanations for fine temporal variations in storm surges that were not addressed in the two aforementioned studies.

In Section \ref{Analysis of meteorological observations}, typhoon tracks affecting the northern coast of Kyushu Island are organized, while patterns leading to significant storm surges along the northern coast of Kyushu Island are identified and classified. Furthermore, observational data on wind, atmospheric pressure, and the characteristics of storm surge anomalies specific to each typhoon’s trajectory pattern are summarized. Section \ref{Numerical simulation} describes the numerical storm surge simulation model used in this study, referring to storm surges during typhoons which are organized in Section \ref{Analysis of meteorological observations}. This section also explains the simulation of storm surges with separate forcings from atmospheric pressure and wind, determining the contributions of those forcings to storm surges. Section \ref{Discussion of storm surge development mechanisms} presents a detailed discussion of storm surge development mechanisms for each trajectory pattern. Finally, Section \ref{Conclusion} consolidates the findings.

\section{Analysis of meteorological observations}\label{Analysis of meteorological observations}

\subsection{Classification of Typhoons}\label{Classification of Typhoons}
\begin{figure}
    \centering
    \includegraphics[width=0.5\linewidth]{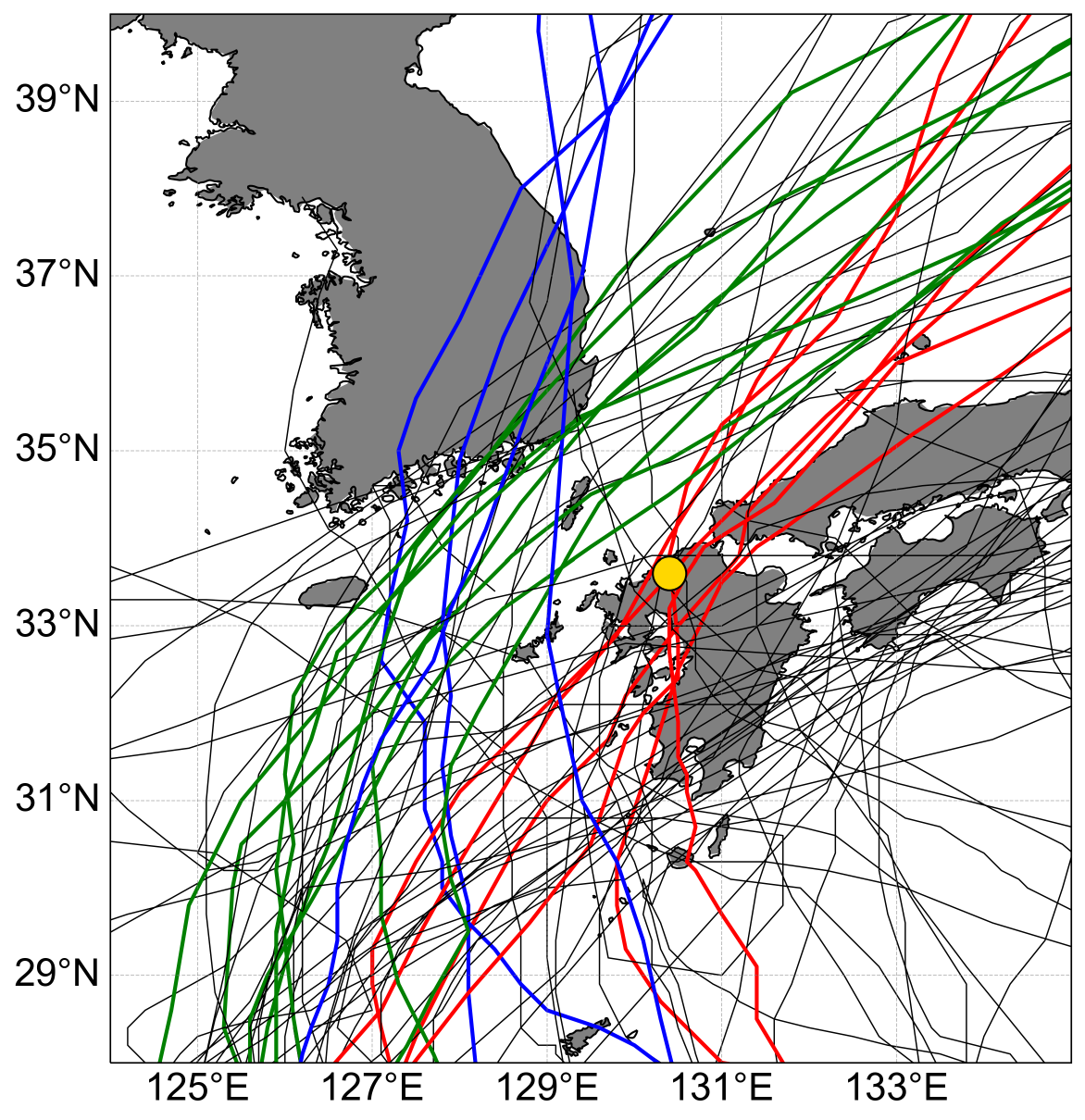}
    \caption{The paths of typhoons passing within a radius of 500~km of Hakata Bay. Colored paths represent the trajectories of typhoons that resulted in a maximum storm surge of 30~cm or more in Hakata Bay. The green paths correspond to northeastward-moving typhoons, the blue paths indicate northward-moving typhoons, and the red paths represent directly passing overhead typhoons.}
    \label{fig:selected_typhoon_course}
\end{figure}

\begin{table}
\centering
\caption{Maximum storm surge anomalies ($\eta_{\mathrm{max}}$) in Hakata Bay tidal gauge station due to selected typhoons.}
\label{table:selected_typhoon_maxssa}
\begin{tabular}{clccll}
\hline
Type                                                                                   & Typhoon name (ID) & \multicolumn{1}{l}{$\eta_{\mathrm{max}}$ (cm)} & Type                                                                                        & Typhoon name (ID) & $\eta_{\mathrm{max}}$ (cm) \\ \hline
\multirow{8}{*}{\begin{tabular}[c]{@{}c@{}}Northeastward \\ moving types\end{tabular}} & MAEMI(T0314)      & 52.95                                          & \multirow{6}{*}{\begin{tabular}[c]{@{}c@{}}Directly passing\\  overhead types\end{tabular}} & BART(T9918)       & \multicolumn{1}{c}{52.10}   \\
                                                                                       & MEGI(T0415)       & 49.63                                          &                                                                                             & CHABA(T0416)      & \multicolumn{1}{c}{63.35}  \\
                                                                                       & DANAS(T1324)      & 33.72                                          &                                                                                             & SONGDA(T0418)     & \multicolumn{1}{c}{86.43}  \\
                                                                                       & CHABA(T1618)      & 36.21                                          &                                                                                             & SHANSHAN(T0613)   & \multicolumn{1}{c}{38.37}  \\
                                                                                       & PRAPIROON(T1807)  & 32.65                                          &                                                                                             & GONI(T1515)       & \multicolumn{1}{c}{46.17}  \\
                                                                                       & KONG-REY(T1825)   & 51.35                                          &                                                                                             & NANMADOL(T2214)   & \multicolumn{1}{c}{37.67}  \\ \cline{4-6} 
                                                                                       & TAPAH(T1917)      & 48.85                                          & \multicolumn{1}{l}{}                                                                        &                   &                            \\
                                                                                       & HINNAMNOR(T2211)  & 65.99                                          & \multicolumn{1}{l}{}                                                                        &                   &                            \\ \cline{1-3}
\multirow{4}{*}{\begin{tabular}[c]{@{}c@{}}Northward \\ moving types\end{tabular}}     & RUSA(T0215)       & 46.63                                          & \multicolumn{1}{l}{}                                                                        &                   &                            \\
                                                                                       & SANBA(T1216)      & 45.42                                          & \multicolumn{1}{l}{}                                                                        &                   &                            \\
                                                                                       & MAYSAK(T2009)     & 56.10                                           & \multicolumn{1}{l}{}                                                                        &                   &                            \\
                                                                                       & HAISHEN(T2010)    & 35.78                                          & \multicolumn{1}{l}{}                                                                        &                   &                            \\ \cline{1-3}
\end{tabular}
\end{table}
In this study, we selected typhoons that caused relatively large storm surges in the target area along the northern coast of Kyushu Island. The criteria for selecting typhoons were as follows: occurred from 1999 to 2022 and passed within a radius of 500~km with Hakata Bay at the center. Storm surge anomalies were calculated by subtracting the astronomical tides publicly available from the Japan Meteorological Agency (\url{https://www.data.jma.go.jp/gmd/kaiyou/db/tide/genbo/genbo.php}) from the observed tidal levels (Note: values before 2011 were calculated based on the harmonic constant table issued by the Japan Coast Guard Hydrographic and Oceanographic Department, available at \url{https://ci.nii.ac.jp/ncid/BN1029405X}).

Figure~\ref{fig:selected_typhoon_course} shows the selected typhoon paths. The colored lines represent typhoon trajectories with a maximum storm surge of 30~cm or more. Typhoons judged to have tidal anomalies that could not be adequately corrected were excluded. The typhoon paths can be classified into the following three trajectory patterns:
\begin{itemize}
    \item Northeastward-moving typhoon (green lines in Figure~\ref{fig:selected_typhoon_course}): A trajectory that passes through the Tsushima Strait from the East China Sea and exits into the Sea of Japan.
    \item Northward-moving typhoon (blue lines in Figure~\ref{fig:selected_typhoon_course}): A trajectory that progresses in a true north direction from off the west coast of Kyushu Island, traverses the western part of the Tsushima Strait, and continues northward even after making landfall on the Korean Peninsula.
    \item Directly passing overhead typhoon (red lines in Figure~\ref{fig:selected_typhoon_course}): A trajectory that advances in a northeast direction from over the East China Sea or along the west coast of Kyushu Island, passing directly over Hakata Bay on the northern coast of Kyushu Island before exiting into the Sea of Japan.
\end{itemize}

Table~\ref{table:selected_typhoon_maxssa} presents the maximum storm surge anomalies caused by selected typhoons in Hakata Bay. The most prevalent trajectory pattern was the northeastward-moving type, followed by a significant occurrence of directly passing overhead typhoons. The least frequent type was the northward-moving type, which accounted for half as many occurrences as the northeastward-moving type. Moreover, directly passing overhead typhoons exhibited the largest maximum storm surge anomalies, averaging 54~cm, followed by the northeastward-moving and northward-moving types at 46~cm. It is noteworthy that, among the typhoons which caused substantial storm surge disasters on the northern coast of Kyushu Island in the past, the typhoon in 1828, named "Seibolt Typhoon," and Typhoon Mireille, which occurred in 1991, are classified as the directly passing overhead type.

\subsubsection{Northeastward-moving types}\label{Northeastward movement}

\begin{figure}
    \centering
    \includegraphics[width=0.5\linewidth]{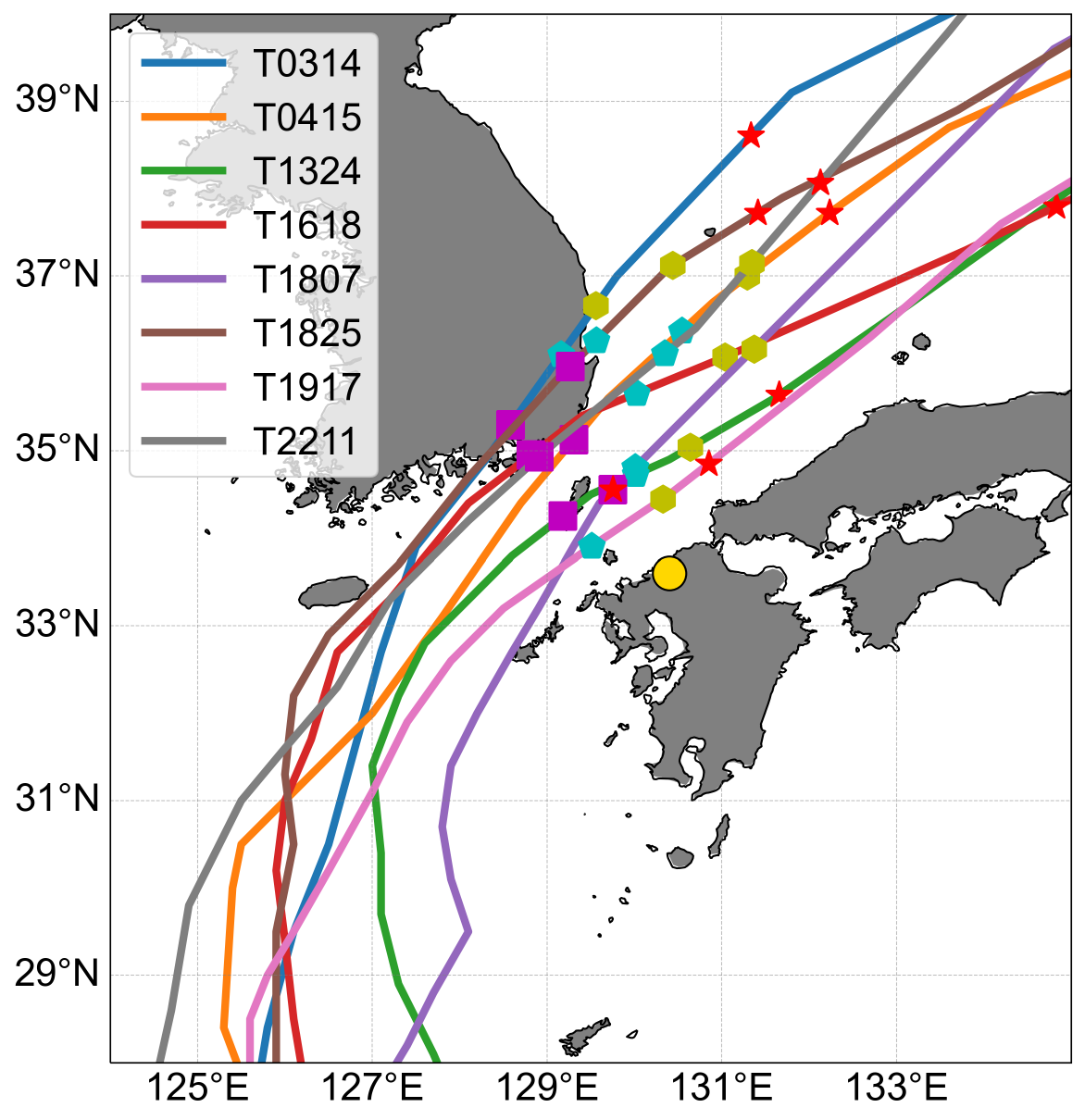}
    \caption{The paths of northeastward-moving types}
    \label{fig:selected_typhoon_course_ne}
\end{figure}
\begin{figure}
    \centering
    \includegraphics[width=\linewidth]{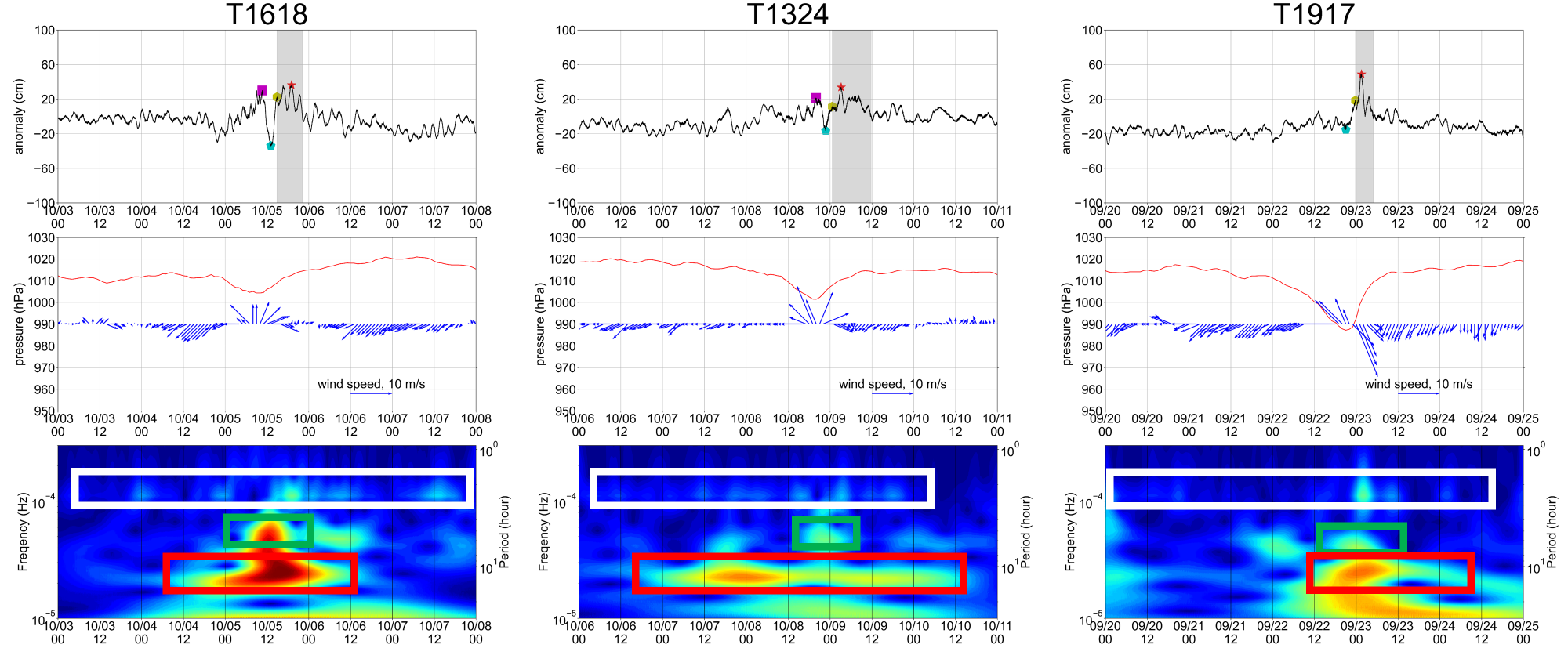}
    \caption{Time-series of storm surge anomalies in Hakata Bay caused by northeastward-moving typhoons (top), along with atmospheric pressure and wind patterns (middle), and continuous wavelet transforms (bottom).}
    \label{fig:obs_timeseries_ne}
\end{figure}
First, we determined the characteristics of storm surges caused by northeastward-moving typhoons. Figure~\ref{fig:selected_typhoon_course_ne} illustrates the trajectories of the typhoons. The yellow dots in Figure~\ref{fig:selected_typhoon_course_ne} represent the location of Hakata Bay.
Furthermore, Figure~\ref{fig:obs_timeseries_ne} displays the time series of storm surge anomalies in Hakata Bay caused by typhoons classified as northeastward-moving (top), along with atmospheric pressure and wind data (middle). The black line in the top panel represents the storm surge anomaly at the Hakata Bay Tidal Gauge Station (\url{https://near-goos1.jodc.go.jp/vpage/search.html}). The blue and red arrows in the middle panel represent wind speed and sea-level pressure, respectively, at the Fukuoka Regional Headquarters  (\url{https://www.jma-net.go.jp/fukuoka/}). The bottom panel shows the continuous wavelet transform of the storm surge anomaly.
The symbols along the storm surge anomaly in Figure~\ref{fig:obs_timeseries_ne} indicate specific events: magenta squares denote the first peak, cyan pentagons represent the minimum peak, yellow hexagons mark the beginning of the second peak, and red stars correspond to the maximum storm surge anomaly. The positions of the typhoons during these events are marked with the same symbols as in Figure~\ref{fig:selected_typhoon_course_ne}.
Additionally, the shaded gray areas in Figure~\ref{fig:obs_timeseries_ne} represent the duration from the onset of the second peak (yellow hexagon) to the point where the storm surge anomaly reached 0~cm, indicating the duration of the elevated surge levels after the second peak.

T1618 intrudes into the Tsushima Strait from the East China Sea, passing along the southern coast of the Korean Peninsula  and entering the Sea of Japan (Figure~\ref{fig:selected_typhoon_course_ne}: T1618). In Hakata Bay, there is a consistent occurrence of approximately 2-hour oscillations. As the typhoon approaches Hakata Bay and atmospheric pressure decreases, the storm surge anomaly gradually rises. When the typhoon makes landfall on the Korean Peninsula, the storm surge anomaly reaches its maximum value, marking the first peak (Figure~\ref{fig:selected_typhoon_course_ne}, Figure~\ref{fig:obs_timeseries_ne}: magenta square). At this point, the wind speed increases significantly as the typhoon approaches, reaching approximately 10~m/s at the closest approach point. The wind direction shifts westward and northward.
After reaching the first peak in the storm surge anomaly, the anomaly rapidly decreases, reaching a minimum peak (Figure~\ref{fig:selected_typhoon_course_ne}, Figure~\ref{fig:obs_timeseries_ne}: cyan pentagon). During this period, the wind speed remains at approximately 10~m/s, similar to the first peak, but the wind direction slightly rotates to the north-northeastward. 
Subsequently, the storm surge anomaly rises again shortly after (Figure~\ref{fig:selected_typhoon_course_ne}, Figure~\ref{fig:obs_timeseries_ne}: yellow hexagon), culminating in the second peak (Figure~\ref{fig:selected_typhoon_course_ne}, Figure~\ref{fig:obs_timeseries_ne}: red star). The second peak is larger than the first, and the elevated storm surge persists for approximately 10 hours (Figure~\ref{fig:obs_timeseries_ne}: shaded gray area). At the time of the second peak, the typhoon intrudes into the Sea of Japan, causing the atmospheric pressure in Hakata Bay to rise to approximately 1,015~hPa, and the wind becomes a northeastward wind of 0-5~m/s. Finally, the storm surge anomaly slowly decreases, the wind speed diminishes, and the wind direction changes from northeastward to southwestward by 180 degrees.

Typhoons T0314, T0415, T1825, and T2211, which take similar paths (Figure~\ref{fig:selected_typhoon_course_ne}), exhibit storm surge anomalies and meteorological characteristics comparable to T1618. The storm surge anomaly experiences its first peak when the typhoon approaches (magenta square), followed by a rapid decrease (cyan pentagon) and a subsequent rise after the typhoon moves away (yellow hexagon), leading to the second peak (red star). This elevated anomaly is sustained for a certain duration (shaded gray area), followed by a gradual decrease after the storm departs. 

Typhoons T1324 and T1917 (Figure~\ref{fig:obs_timeseries_ne}: center, right) take the path of entering the East China Sea, passing through the Tsushima Strait, closer to the northern coast of Kyushu Island, and then moving into the Sea of Japan (Figure~\ref{fig:selected_typhoon_course_ne}: T1324 and T1917). T1324 exhibits a smaller peak size, and T1917 shows only one peak in the storm surge anomaly (Figure~\ref{fig:obs_timeseries_ne}: red star). Although not shown in the figure,  T1807, which takes a similar path, has a smaller peak size, showing only one peak during the closest approach to the typhoon, reaching its maximum value.
Therefore, among the northeast-moving typhoons passing through the Tsushima Strait, those that pass closer to the southern coast of the Korean Peninsula  exhibit a behavior in the storm surge anomaly in Hakata Bay with two peaks. On the other hand, when typhoons pass closer to the northern coast of Kyushu Island, their behavior changes, and only one peak is observed instead of two.

Next, to examine the components of the oscillations present in the storm surge anomaly in Hakata Bay, a spectral analysis using a continuous wavelet transform (\cite{Lee2019}) was conducted. The continuous wavelet transform is a method for frequency analysis along the time axis, involving two operations: translation (shift) and dilation/contraction (scale) of a localized wave called the mother wavelet. By performing these operations and observing their similarity to the original waveform, the continuous wavelet transform provides insights into the frequency content. The mother wavelet is denoted as 
\(\psi(t)\), the wavelet at scale \(a\) and shift \(b\) is given by:
\begin{equation}
    \psi_{a, b} (t)=\frac{1}{\sqrt{a}}\psi\left( \frac{t-b}{a} \right)
\end{equation}
When \(a\) increases, the mother wavelet is stretched, making it more suitable for analyzing a long wave. Increasing \(b\) results in a parallel shift in the positive direction along the time axis. For the mother wavelet defined by scale \(a\) and shift \(b\), denoted by \(\psi_{a, b} (t)\), the continuous wavelet transform is expressed as follows:
\begin{equation}
    W_{\psi_{a, b}}[x(t)]=\int_{-\infty}^{\infty}x(t)\psi_{a, b}^{*} (t)dt
\end{equation}
Where \(x(t)\) is the target wave and \(*\) denotes the complex conjugate. By continuously varying the scale \(a\) and shift \(b\), the continuous wavelet transform maps onto a two-dimensional plane of \((a, b)\), enabling frequency analysis in the time domain. In this study, the following complex Morlet wavelet is used as the mother wavelet:
\begin{equation}
    \psi(t) = \frac{1}{\sqrt{\pi B}} \exp\left(-\frac{t^2}{B}\right)\exp(2\pi Cjt)
\end{equation}
Where \(j\) is the imaginary unit, \(B\) is the bandwidth, and \(C\) is the center frequency. In this study, \(B=1.5\) and \(C=1.0\).

The lower panel of Figure~\ref{fig:obs_timeseries_ne} shows the wavelet transform of the storm surge anomaly in Hakata Bay caused by the northeast-moving typhoon. For all northeast-moving typhoons, including T1618, T1324, and T1917, oscillations with a period of 2-3 hours were detected throughout the entire period, from before the typhoon approached to after its passage (Figure~\ref{fig:obs_timeseries_ne} lower panel: white box). This oscillatory component exhibits a period very close to the natural period of Hakata Bay (\cite{Yamashiro2016}), indicating that bay oscillations occur before the typhoon's approach and persist continuously afterward.

Moreover, in T1618, oscillations with a period of 7 hours were also detected (Figure~\ref{fig:obs_timeseries_ne} lower panel, highlighted in green). The time at which the 7-hour oscillation reaches its maximum coincides with the occurrence of the minimum peak in the storm surge anomaly. As the surge anomaly exhibits a second peak, the amplitude of the 7-hour oscillation diminishes, becoming nearly invisible. Although not shown in the figure, typhoons with paths similar to T1618, such as T0314, T0415, T1825, and T2211, also exhibited components of the 7-hour oscillation. Conversely, typhoons passing near the northern coast of Kyushu Island, such as T1324, T1807, T1825, and T1917, despite also showing a 7-hour wave, exhibited a smaller amplitude.

Furthermore, the component of the 10-hour oscillation was detected in all northeastward-moving typhoons (Figure~\ref{fig:obs_timeseries_ne} red box). This oscillation component was present more than two days before the typhoon's closest approach to Hakata Bay, and its amplitude peaked at the time of the second peak. Subsequently, the amplitude gradually decreased but persisted for more than two days. 
As oscillations with a period of 7 and 10 hours exhibit an increase in amplitude after the closest approach, understanding their mechanisms and behavior plays an important role in unraveling storm surge development mechanisms on the northern coast of Kyushu Island. A detailed discussion of these mechanisms is provided in Section \ref{Discussion of storm surge development mechanisms}.
\clearpage

\subsubsection{Northward-moving types}\label{Northward movement}

\begin{figure}
    \centering
    \includegraphics[width=0.5\linewidth]{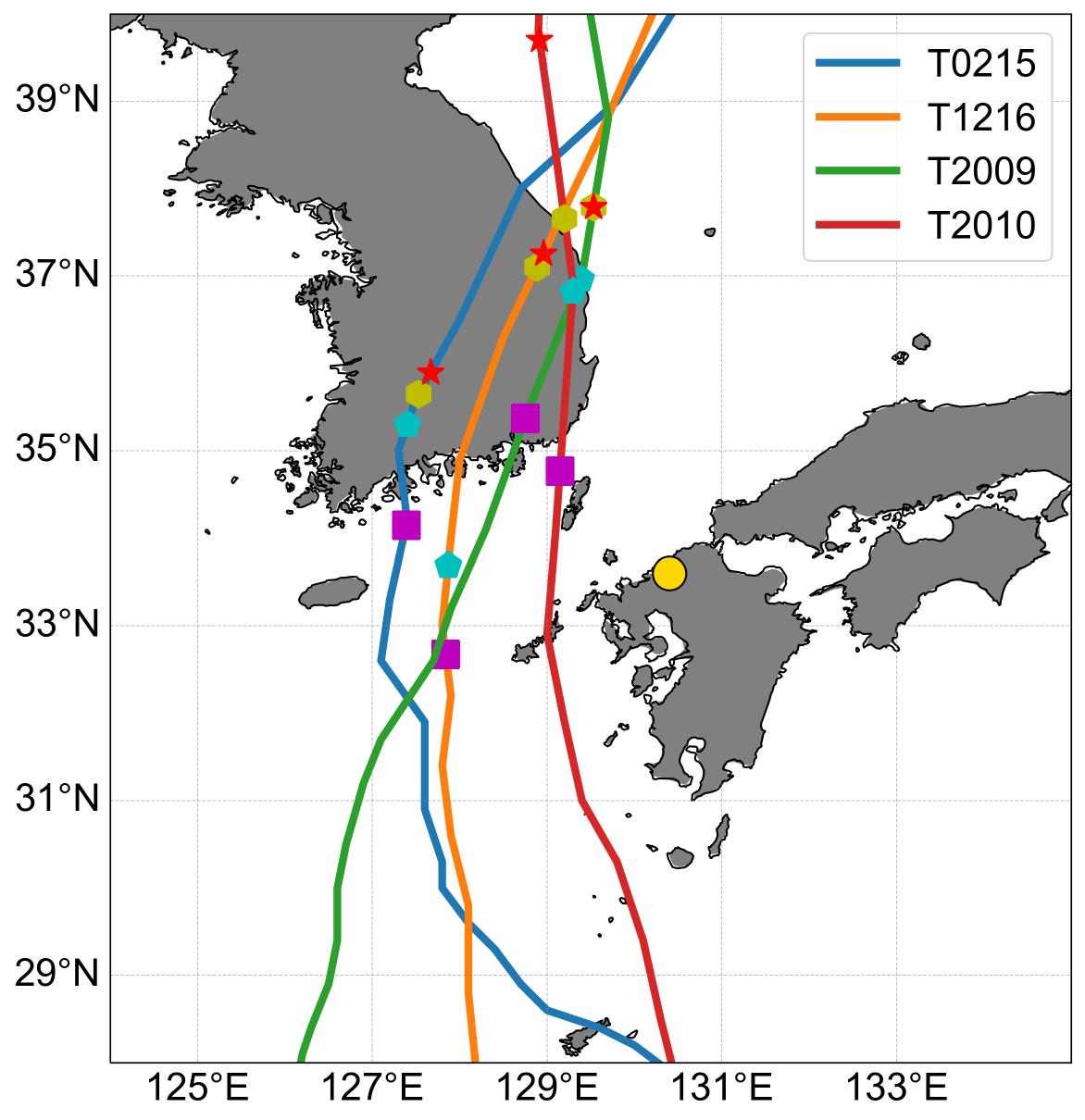}
    \caption{The paths of northward-moving types}
    \label{fig:selected_typhoon_course_n}
\end{figure}
\begin{figure}
\includegraphics[width=\linewidth]{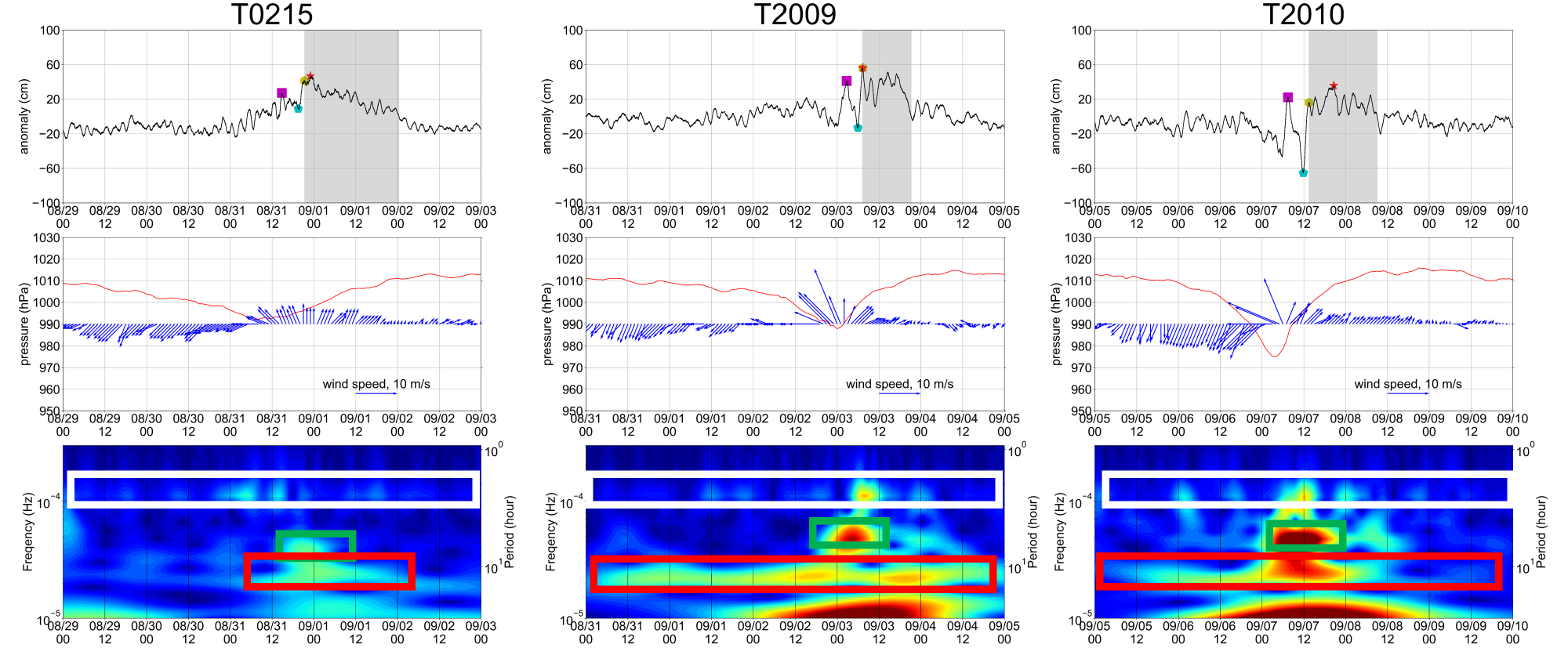}
\caption{Time-series of storm surge anomalies in Hakata Bay caused by northward-moving typhoons (top), along with atmospheric pressure and wind patterns (middle), and continuous wavelet transforms (bottom).}
\label{fig:obs_timeseries_n}
\end{figure}

Figure~\ref{fig:selected_typhoon_course_n} depicts the paths of the northward-moving typhoons. Figure~\ref{fig:obs_timeseries_n} illustrates the storm surge anomalies, atmospheric pressure, and wind data for the northward-moving typhoons (T0215, T2009, and T2010) in Hakata Bay. The surge anomalies, wind, and atmospheric pressure data are obtained from the same observation site shown in Figure~\ref{fig:obs_timeseries_ne}, with the meanings of the lines and symbols consistent with those in Figure~\ref{fig:selected_typhoon_course_ne} and Figure~\ref{fig:obs_timeseries_ne}.

As T0215 approaches, the storm surge anomaly gradually increases. Upon approaching the closest typhoon, it reaches its first peak (Figure~\ref{fig:selected_typhoon_course_n}, Figure~\ref{fig:obs_timeseries_n}: magenta square). Subsequently, the surge anomaly decreases (Figure~\ref{fig:selected_typhoon_course_n}, Figure~\ref{fig:obs_timeseries_n}: cyan pentagon), then rapidly increases (Figure~\ref{fig:selected_typhoon_course_n}, Figure~\ref{fig:obs_timeseries_n}: yellow hexagon), reaching the second peak (Figure~\ref{fig:selected_typhoon_course_n}, Figure~\ref{fig:obs_timeseries_n}: red star). Before the typhoon's arrival, a northward wind with a speed of 5~m/s blows, changing northwestward as the typhoon approaches. In the post-closest approach, a northeastward wind with a speed of 5~m/s continues for 24 hours. After the second peak, a high anomaly is maintained for 24 hours (Figure~\ref{fig:obs_timeseries_n}: gray hatching). T1216 exhibited a similar surge anomaly behavior to T0215, although this is not shown in Figure~\ref{fig:obs_timeseries_n}.

The storm surge anomaly of T2009 exhibited behavior similar to that of northeastward-moving typhoons passing along the southern coast of the Korean Peninsula until the second peak (Figure~\ref{fig:selected_typhoon_course_n}, Figure~\ref{fig:obs_timeseries_n}: red star). Upon the typhoon's approach, it experiences the first peak (Figure~\ref{fig:selected_typhoon_course_n}, Figure~\ref{fig:obs_timeseries_n}: magenta squares), followed by a rapid decrease in the surge anomaly (Figure~\ref{fig:selected_typhoon_course_n}, Figure~\ref{fig:obs_timeseries_n}: cyan pentagons). After the typhoon moves away, the anomaly rises again (Figure~\ref{fig:selected_typhoon_course_n}, Figure~\ref{fig:obs_timeseries_n}: yellow hexagon) and reaches a second peak (Figure~\ref{fig:selected_typhoon_course_n}, Figure~\ref{fig:obs_timeseries_n}: red star). However, owing to the persistent northeastward wind after the second peak, a high anomaly is maintained for approximately 12 hours (Figure~\ref{fig:obs_timeseries_n}: gray hatching). The presence of Ekman transport caused by parallel northeastward winds along the coast may explain the maintenance of a high anomaly for 12 hours. T2010, following a path similar to T2009, maintains a high anomaly for 20 hours after its second peak.

Thus, for northward-moving typhoons, the northeastward wind persists even after the typhoon passes through the Tsushima Strait, and there is a tendency for a high storm surge anomaly to be maintained for an extended period after the second peak, driven by Ekman transport. If the high anomaly coincides with high tide, there is a risk of significantly elevated water levels and flooding. 
In summary, northward-moving typhoons have a higher likelihood of overlapping high-surge anomalies with high tides than do northeastward-moving typhoons.

Next, we conducted the spectral analysis using a continuous wavelet transform. The lower panel of Figure~\ref{fig:obs_timeseries_n} shows the wavelet transform of storm surge anomalies in Hakata Bay caused by northward-moving typhoons. For all northward-moving typhoons, harbor oscillations in Hakata Bay, around 2-3 hours, were observed (Figure~\ref{fig:obs_timeseries_n}: white box). In addition, oscillations with a period of 10 hours (Figure~\ref{fig:obs_timeseries_n}: red box) were identified in both T0215 and T2009.
Conversely, 7~hour periodic oscillations (Figure~\ref{fig:obs_timeseries_n}: green box) were observed in T2009 but with smaller amplitudes in T0215. Compared to other northward-moving typhoons, T2010 exhibited a spectrum similar to T2009, whereas T1216 showed a spectrum resembling T0215. T0215 and T1216 followed paths passing to the west of 128°E in the Tsushima Strait, reaching the southern coast of the Korean Peninsula. It is believed that this slight variation in the trajectory prevented the abrupt 7~hour periodic oscillations observed in T2009 and T2010, as well as in northeastward-moving typhoons passing near the southern coast of the Korean Peninsula, such as T0215.
\clearpage

\subsubsection{Directly passing overhead types}\label{straight up through}

\begin{figure}
    \centering
    \includegraphics[width=0.5\linewidth]{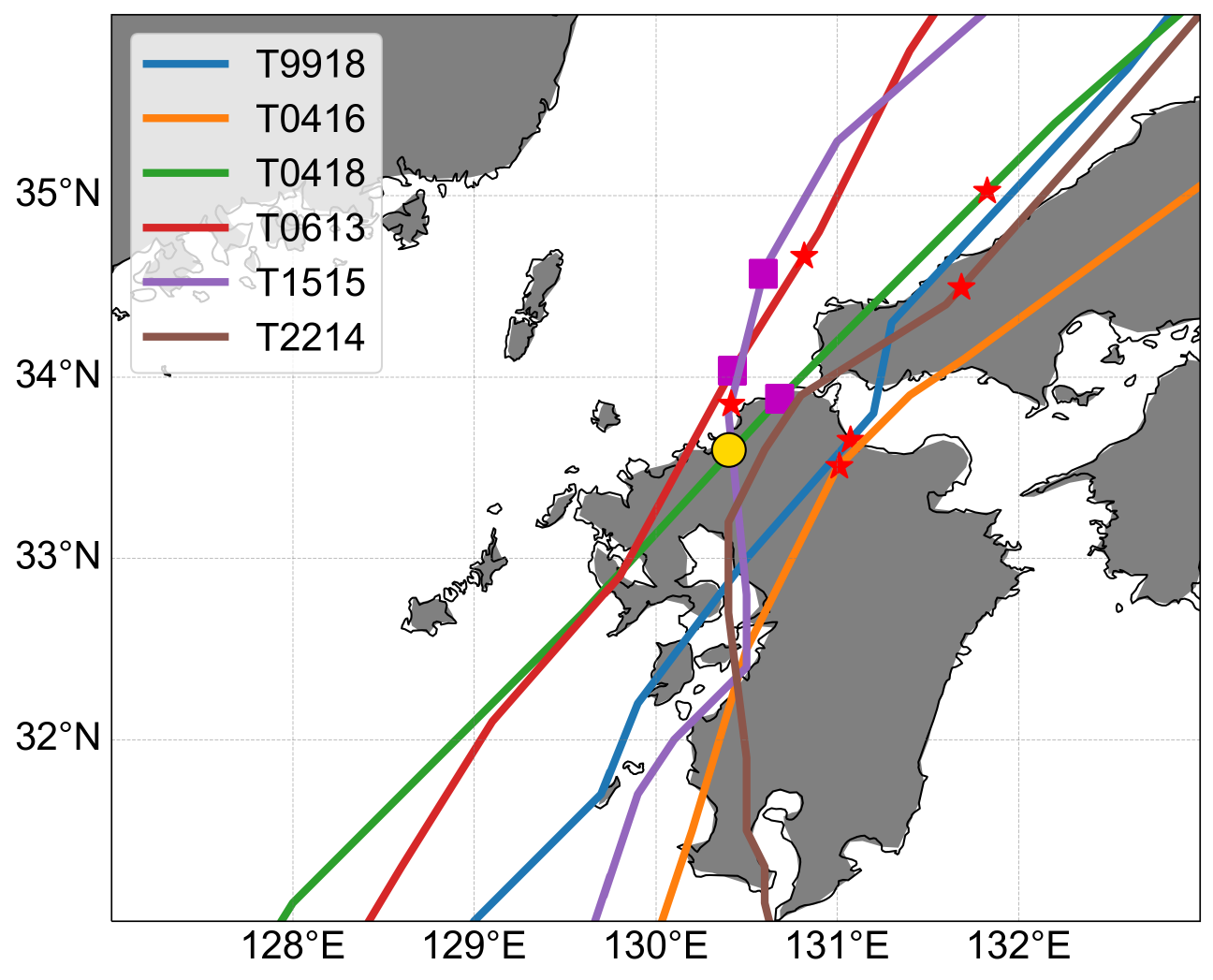}
    \caption{The paths of directly passing overhead types}
    \label{fig:selected_typhoon_course_st}
\end{figure}
\begin{figure}
\includegraphics[width=\linewidth]{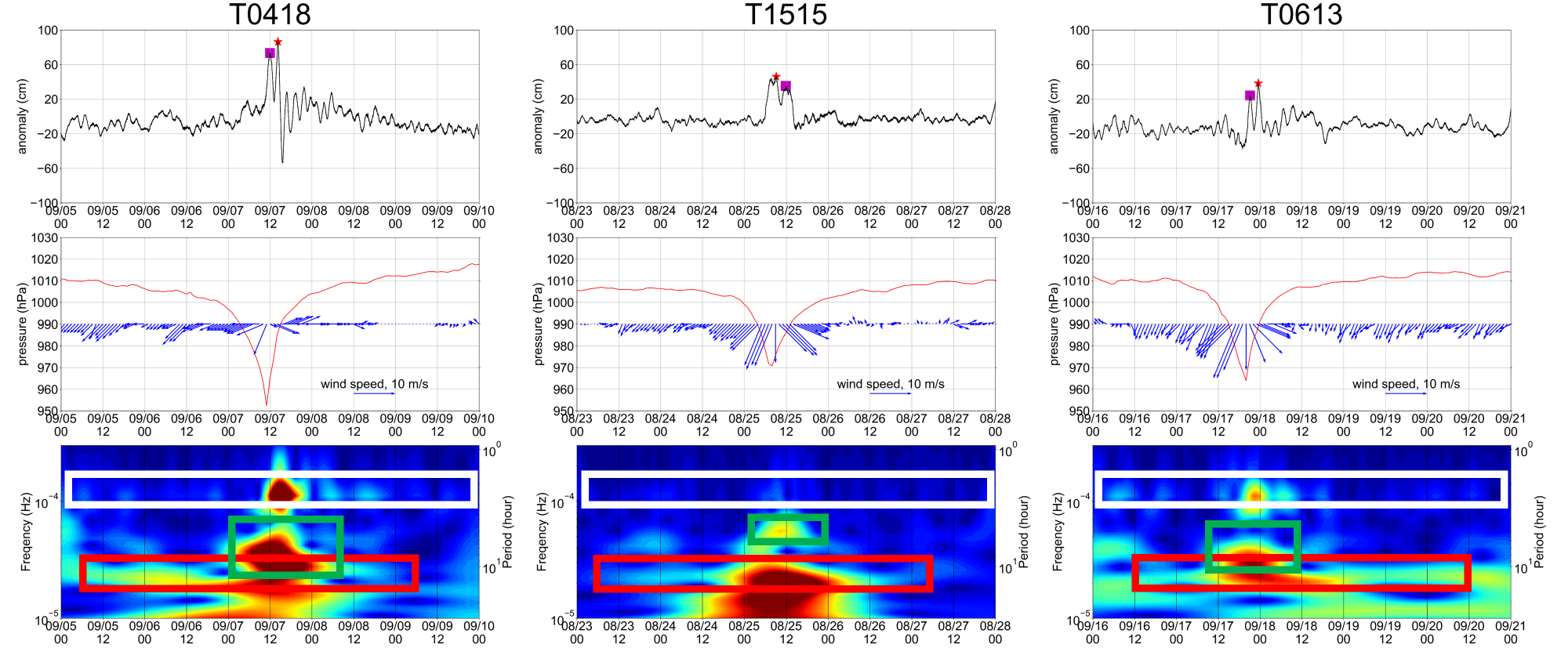}
\caption{Time-series of storm surge anomalies in Hakata Bay caused by directly passing overhead typhoons (top), along with atmospheric pressure and wind patterns (middle), and continuous wavelet transforms (bottom).}
\label{fig:obs_timeseries_st}
\end{figure}
Figure~\ref{fig:selected_typhoon_course_st} presents the individual paths of typhoons of the direct-passing overhead type, whereas Figure~\ref{fig:obs_timeseries_st} illustrates the storm surge anomalies, meteorological observations, and continuous wavelet transforms for this type, including T0418, T1515, and T0613. The symbols for storm surge anomalies indicate larger peaks with red stars and smaller peaks with magenta squares when there are two peaks.

When typhoons of the direct-passing overhead type approach Hakata Bay, the atmospheric pressure rapidly decreases to approximately 960-970~hPa, accompanied by southward winds. Subsequently, shortly after or slightly delayed from the closest approach, a single or double peak in the storm surge anomalies occurs. T9918 and T0416 exhibit a single peak, whereas T0418, T0613, and T1515 exhibit two peaks. Following this, storm surge anomalies quickly decrease and the atmospheric pressure rises to approximately 1,010~hPa. In addition, the wind shifts from southward to southeastward, rapidly decreasing in speed. However, T2214, which is not shown in Figure~\ref{fig:obs_timeseries_st}, gradually peaks 10 hours after the closest approach.
Typhoons T9918 and T0416, with a single peak in storm surge anomalies, pass through locations far from Hakata Bay. By contrast, typhoons T0418, T0613, and T1515, exhibiting two peaks, pass through locations relatively close to Hakata Bay. Typhoons approaching Hakata Bay have a higher likelihood of exhibiting the two peaks.
Therefore, typhoons of the direct-passing type, unlike those of the northeastward or northward type, tend to rapidly increase storm surge anomalies to a significant extent in a short period during the closest approach.

The lower panel of Figure~\ref{fig:obs_timeseries_st} shows the wavelet transform of the storm surge anomalies in Hakata Bay caused by direct passing-type typhoons. The 2-hour natural period of Hakata Bay (white box) is observed in T0418, with a significant amplitude during the closest approach to the typhoon, diminishing as the typhoon moves away. A component with a 7-hour range (green frame) is also identified. This component corresponds to variations in atmospheric pressure and reflects an increase in storm surge anomalies due to the inverted barometer effect. Furthermore, a 10-hour periodic oscillation (red frame) is observed before the typhoon's approach.
In T1515, the 2-hour periodic oscillation is discernible, although with a small amplitude. During the closest approach of the typhoon, a 7-hour periodic oscillation emerges, correlating with fluctuations in atmospheric pressure, thus indicating the influence of the inverted barometer effect. Additionally, a 10~hour periodic oscillation component is evident.
\clearpage

\subsection{Analysis of storm surge anomalies along the Tsushima Strait Coast}\label{Analysis of storm surge anomalies along the Tsushima Strait Coast}

In Subsection \ref{Analysis of meteorological observations}, components with a 2~hour natural period, 7~hour periodicity, and a 10~hour periodicity were identified. However, the spatial scales of the 7~hour and 10~hour periodic waves were unclear. Therefore, a comparative analysis of storm surge anomalies at multiple observation points along the Tsushima Strait was conducted to investigate the spatial scales of the oscillations.
\begin{figure}
    \centering
    \includegraphics[width=0.5\linewidth]{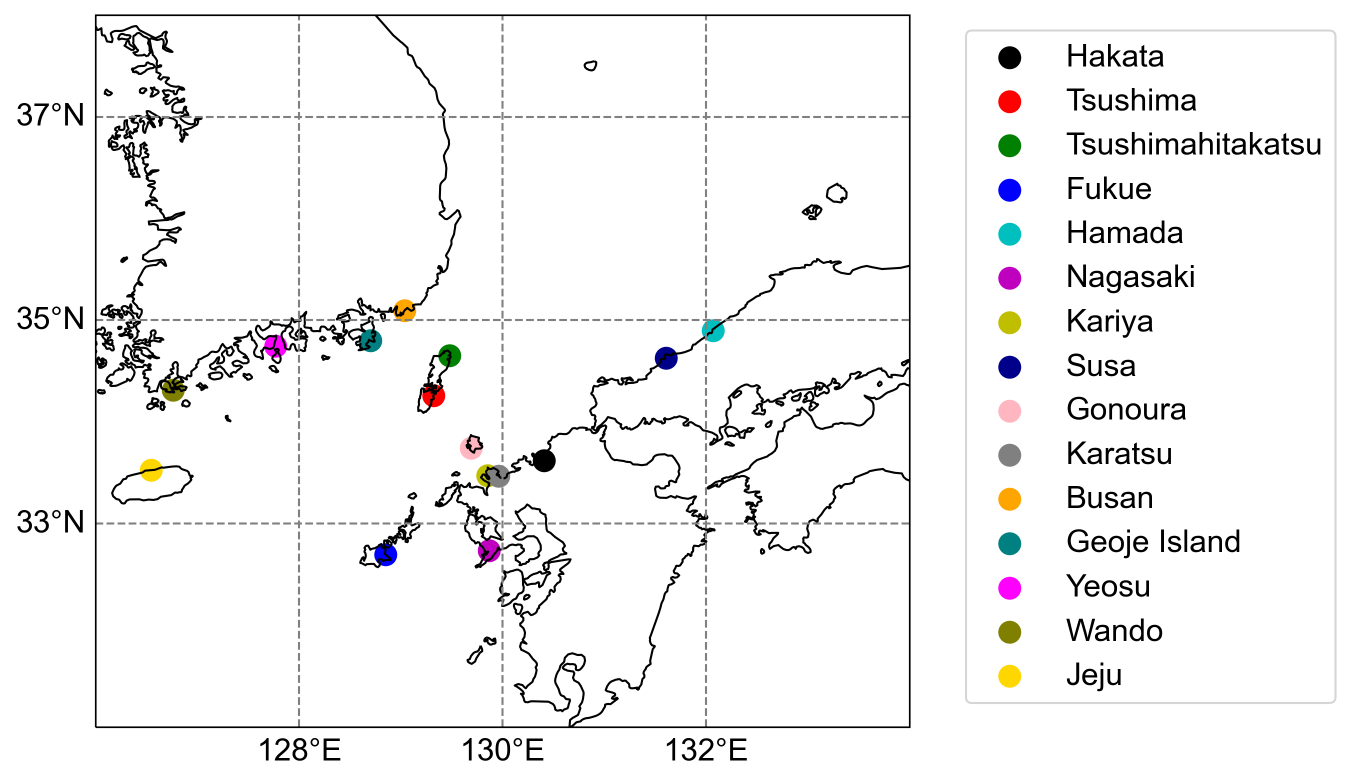}
    \caption{Location of tidal gauge stations along the Tsushima Strait coast.}
    \label{fig:obs_map}
\end{figure}

The locations of tidal gauge stations used in the analysis are depicted in Figure~\ref{fig:obs_map}. In addition to Hakata Bay, data were obtained from 14 other stations. For tidal levels, data about Tsushima, Tsushimahitakatsu, Nagasaki, Hamada, Gonoura, Karatsu, Fukue, Kariya, and Susa were acquired from the Japan Oceanographic Data Center of the Japan Meteorological Agency (\url{https://www.jodc.go.jp/jodcweb/index_j.html}). Storm surge anomalies were calculated by subtracting the astronomical tides obtained from the Japan Meteorological Agency (JMA)'s tidal tables (\url{https://www.data.jma.go.jp/gmd/kaiyou/db/tide/suisan/index.php}). The tidal anomalies for Busan, Geoje Island, Yeosu, Wando, and Jeju in the Korean Peninsula were acquired from the Korea Hydrographic and Oceanographic Agency (\url{https://www.khoa.go.kr/eng/Main.do}). All data were collected at 1-hour intervals. For continuous gaps within four hours, third-order spline interpolation was performed, and the analysis focused on T1618, where 7~hour and 10~hour periodic oscillations were identified.

\begin{figure}
    \centering
    \includegraphics[width=\linewidth]{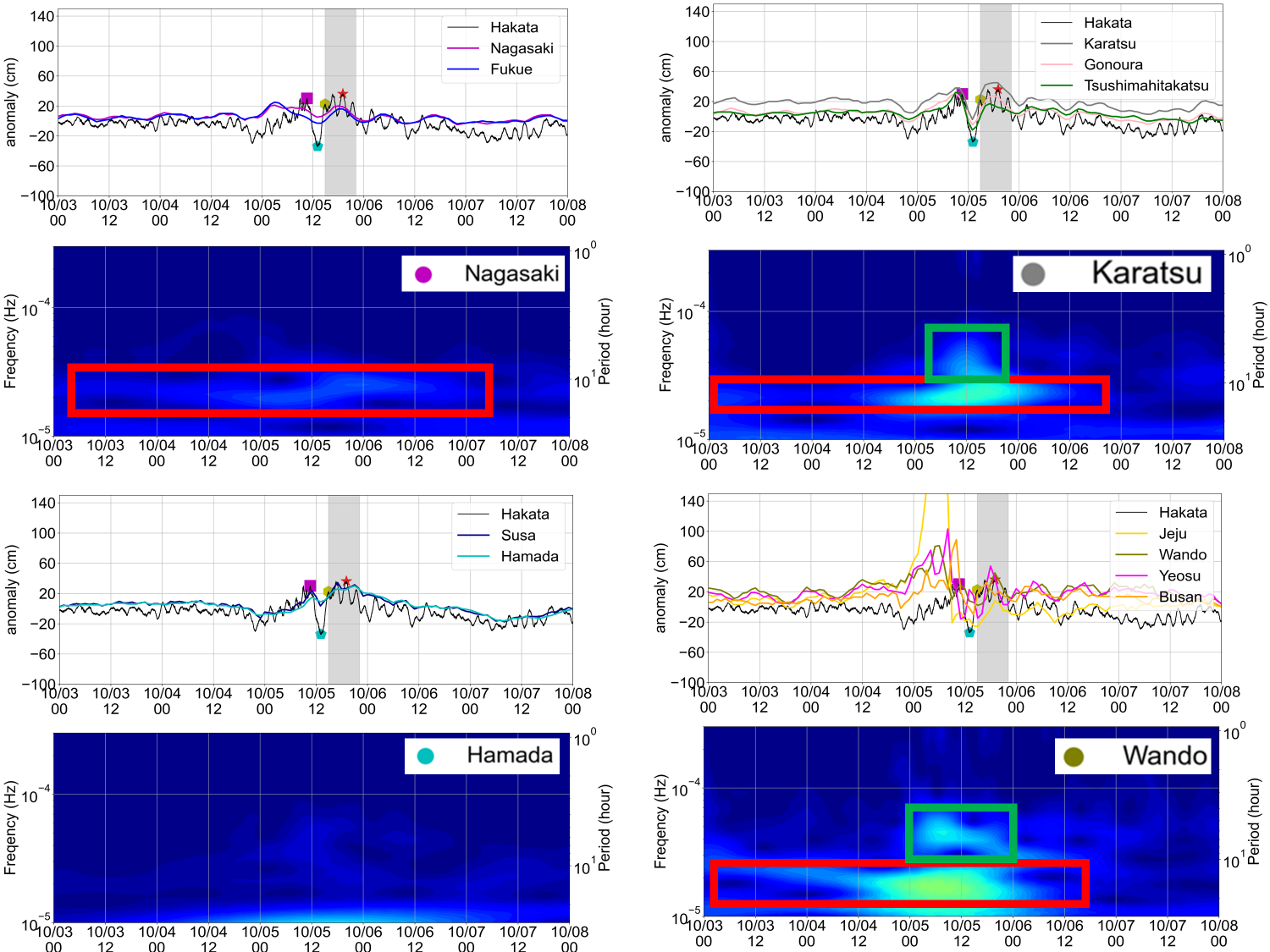}
    \caption{Storm surge anomalies and wavelet transform for each observation point during Typhoon T1618.}
    \label{fig:obs_wavelet_all_stations_T1618}
\end{figure}
\begin{figure}
    \centering
    \includegraphics[width=0.6\linewidth]{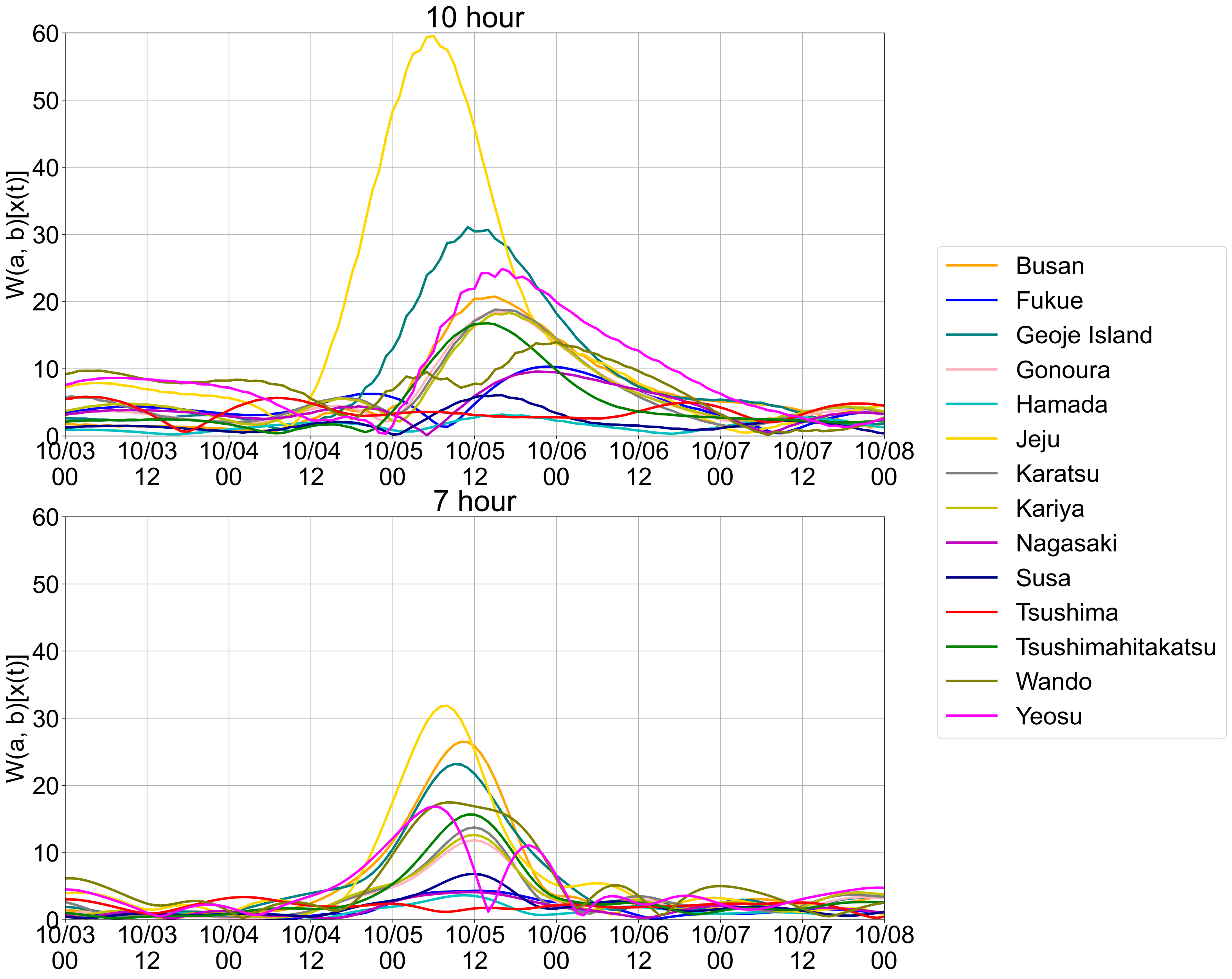}
    \caption{Components of the 10~hour (upper panel) and 7~hour period (lower panel) for each tidal gauge station during Typhoon T1618.}
    \label{fig:obs_wavelet_all_stations_T1618_10h7h}
\end{figure}

Figure~\ref{fig:obs_wavelet_all_stations_T1618} illustrates the storm surge anomalies and wavelet transforms at each observation point during T1618. Firstly, Nagasaki, Fukue, and Hakata are compared (Figure~\ref{fig:obs_wavelet_all_stations_T1618}: top left). Storm surge anomalies in Hakata Bay reach their first peak earlier than the recorded time of the first peak on October 5th at 10:30. Nagasaki and Fukue, on the other hand, reach their first peak at 03:00 on October 5th, with a magnitude of approximately 20~cm. Subsequently, the storm surge anomalies at Nagasaki and Fukue gradually decrease, reaching their minimum peaks at approximately 13:00 on October 5th. At this time, the storm surge anomaly in Hakata Bay reaches its minimum. On the other hand, the minimum peaks at Nagasaki and Fukue are not as distinct as those at Hakata Bay. Following the minimum peaks, the storm surge anomalies at Nagasaki and Fukue slowly increase compared to Hakata, reaching their second peaks simultaneously with Hakata Bay at 19:00 on October 5th. Subsequently, storm surge anomalies gradually decrease with accompanying oscillations. Examining the spectrum for Nagasaki, the 10~hour periodic oscillation consistently dominates (highlighted in red), with the amplitude reaching its maximum around the time of the first peak in the storm surge anomalies. Fukue exhibits a spectrum similar to that of Nagasaki, although this is not shown.

Observation points located near the central region of Tsushima Strait and the northern coast of Kyushu Island, such as Karatsu, Hamada, Gonoura, and TsushimahitaKatsu, exhibit storm surge anomalies very similar to those in Hakata Bay, excluding the bay's 2~hour natural oscillation (Figure~\ref{fig:obs_wavelet_all_stations_T1618}: top right). The spectrum also closely resembles that of Hakata Bay, with noticeable 10~hour periodic oscillations (highlighted in red) and 7~hour periodic oscillations (highlighted in green).

Conversely, stations such as Susa and Hamada exhibit storm surge anomalies, with the first and second peaks occurring simultaneously and showing magnitudes similar to those of Hakata Bay. After the first peak, the storm surge anomalies do not decrease significantly (Figure~\ref{fig:obs_wavelet_all_stations_T1618}, bottom left). Subsequently, storm surge anomalies gradually decrease. Examining the Hamada, 10~hour periodic oscillations are not evident.

Finally, a comparison was made between stations on the south coast of the Korean Peninsula, including Busan, Yeosu, Wando, and Jeju. Jeju experiences its first peak early, with a storm surge anomaly reaching 202~cm at 4:00 on October 5th. Wando follows one hour later, reaching its first peak at 5:00 on October 5th, followed by Yeosu at 8:00, and Busan at 10:00 on the same day. Along the southern coast of the Korean Peninsula, peaks in storm surge anomalies occur sequentially as the typhoon approaches, with each observation point experiencing its first peak earlier than Hakata Bay. Additionally, the magnitude of the storm surge anomalies is exceptionally large. Because northwest winds prevail when a typhoon approaches the south coast, it is believed that significant storm surges are induced by wind setup there. Examining the spectrum for Wando, the 10~hour periodic oscillation is dominant (red frame) and reaches its maximum during the typhoon's closest approach.

The lower panel of Figure~\ref{fig:obs_wavelet_all_stations_T1618_10h7h} shows the 7~hour periodic components at each observation point based on T1618. The colors of each line correspond to the legend in Figure~\ref{fig:obs_wavelet_all_stations_T1618}. The 7~hour periodic components correspond to the magnitude of the minimum peak, while Busan, Jeju, Geoje Island, Yeosu, and Wando exhibit large values. Similarly, the northern coast of Kyushu Island, including Tsushimahitakatsu, Karatsu, Kariya, and Gonoura, also records relatively large values. In contrast, Susa, Hamada, Nagasaki, Fukue, and Tsushima show almost no detection of the 7~hour periodic waves. Therefore, it is found that the 7~hour periodic waves occur in the Tsushima Strait, but within limited areas.

The upper panel of Figure~\ref{fig:obs_wavelet_all_stations_T1618_10h7h} shows the 10~hour periodic components at each observation point based on T1618. In contrast with the 7~hour periodic waves, the 10~hour periodic waves are observed across the Tsushima Strait, except for two points near the Sea of Japan (Susa and Hamada). These waves are detected starting on October 5th at 0:00, when the typhoon enters the Tsushima Strait, and persist for approximately two days. Although there is no significant phase difference in the oscillations between the north coast of Kyushu and the south coast of the Korean Peninsula, the south coast of the Korean Peninsula exhibits larger components than the north coast of Kyushu. Additionally, even on the northwest coast of Kyushu Island, a small component of the 10-hour periodic oscillation is observed. Therefore, the 10-hour periodic oscillations appear to be a coherent phenomenon occurring throughout the Tsushima Strait, with the vibrations having nodes along the boundary between the Sea of Japan and the Tsushima Strait. The reasons for the 10~periodic oscillations adopting such spatial scales are discussed in Section \ref{Discussion of storm surge development mechanisms}.
\clearpage

\section{Numerical simulation}\label{Numerical simulation}
\subsection{Summary of FVCOM}

The storm surge model utilizes the Finite-Volume Community Ocean Model version 3.2 (FVCOM \cite{Chen2003}). The governing equations consist of momentum and continuity equations, as follows:
\begin{align}
    \frac{\partial u}{\partial t} + u\frac{\partial u}{\partial x} + v\frac{\partial u}{\partial y} + w\frac{\partial u}{\partial z} - fv &=  -\frac{1}{\rho_0} \frac{\partial (p_H + p_a)}{\partial x} + \frac{\partial}{\partial z} \left(K_m \frac{\partial u}{\partial z}\right) + F_u \label{u} \\
    \frac{\partial v}{\partial t} + u\frac{\partial v}{\partial x} + v\frac{\partial v}{\partial y} + w\frac{\partial v}{\partial z} + fu &= -\frac{1}{\rho_0} \frac{\partial (p_H + p_a)}{\partial y} + \frac{\partial}{\partial z} \left(K_m \frac{\partial v}{\partial z}\right) + F_v \label{v} \\
    \frac{\partial w}{\partial t} + u\frac{\partial w}{\partial x} + v\frac{\partial w}{\partial y} + w\frac{\partial w}{\partial z} &=  \frac{\partial}{\partial z} \left(K_m \frac{\partial w}{\partial z}\right) + F_w \label{w} \\  
    \frac{\partial u}{\partial x} + \frac{\partial v}{\partial y} + \frac{\partial w}{\partial z} &=  0 \label{c}
\end{align}
where, \(x\), \(y\) and \(z\) represent the horizontal and vertical components in the Cartesian coordinate system, while \(u\), \(v\), and \(w\) represent the velocity components in the \(x\), \(y\), and \(z\) directions, respectively. \(p_a\) and \(p_H\) denote the atmospheric pressure at sea level and hydrostatic pressure, while \(f\) is the Coriolis parameter, \(g\) is the acceleration due to gravity, \(K_m\) is the vertical eddy viscosity coefficient, \(\rho_0\) is the density of seawater, and \(F_u\),\(F_v\), and \(F_w\) represent the horizontal momentum diffusion term and vertical momentum diffusion term, respectively.

The boundary conditions at the sea surface are as follows:
\begin{align}
    K_m \left(\frac{\partial u}{\partial z}, \frac{\partial v}{\partial z} \right) &= \frac{1}{\rho_0} (\tau_{sx}, \tau_{sy}) \label{bc_s}
    \\w &= \frac{\partial \eta}{\partial t} + u \frac{\partial \eta}{\partial x} + v \frac{\partial \eta}{\partial y}
\end{align}
where $\eta$ is the height of the free surface. $\tau_{sx}$ and $\tau_{sy}$ represent \(x\)- and 
\(y\)-components of the sea surface shear stress and are expressed by the following equations:
\begin{align}
    (\tau_{sx}, \tau_{sy}) &= \rho_a C_s U (U_{10}, V_{10}) \label{taus}
    \\ U &\equiv \sqrt{U_{10}^2 + V_{10}^2}
\end{align}
where, $U_{10}$ and $V_{10}$ are \(x\) and 
\(y\)-components of the wind speed at 10m above the sea surface, and $\rho_a$ is the density of the atmosphere. $C_s$ is the drag coefficient of the sea surface, which is calculated using the following equation by referring to the formula in \cite{Large1981}:
\begin{equation}
    C_s =
      \left\{
      \begin{array}{ll}
      1.20 \times 10^{-3}  & \quad \text{if} \quad 0 \le U < 11\text{m/s} \\
      (0.49 + 0.065U) \times 10^{-3} & \quad \text{if} \quad 11\text{m/s} \le U
      \end{array}
      \right.
\end{equation}

The boundary conditions at the seabed are as follows:
\begin{align}
    K_m \left(\frac{\partial u}{\partial z}, \frac{\partial v}{\partial z} \right) &= \frac{1}{\rho_0} (\tau_{bx}, \tau_{by}) \label{bc_b}\\
    w &= - u \frac{\partial H}{\partial x} - v \frac{\partial H}{\partial y}
\end{align}
where, $H$ denotes water depth. $\tau_{bx}$ and $\tau_{by}$ represent \(x\)- and 
\(y\)-components of the seabed shear stress and are expressed by the following equations:
\begin{align}
    (\tau_{bx}, \tau_{by}) = \rho_0 C_b \sqrt{u^2 + v^2} (u, v) \label{taub}
\end{align}

The bottom friction coefficient $C_b$ is calculated using the following equation:
\begin{align}
    C_b = \text{max} \left(\kappa^2/\text{ln}\left( \frac{z_a}{z_0} \right)^2, 0.0025 \right) \label{Cb}
\end{align}
where $\kappa(=0.4)$ is the Karman constant, \(z_0\) is the roughness length, and \(z_a\) is the height above the seabed. The initial water level is assumed to be constant at mean sea level, and both tides and waves are not considered.
\begin{figure}
    \centering
    \includegraphics[width=\linewidth]{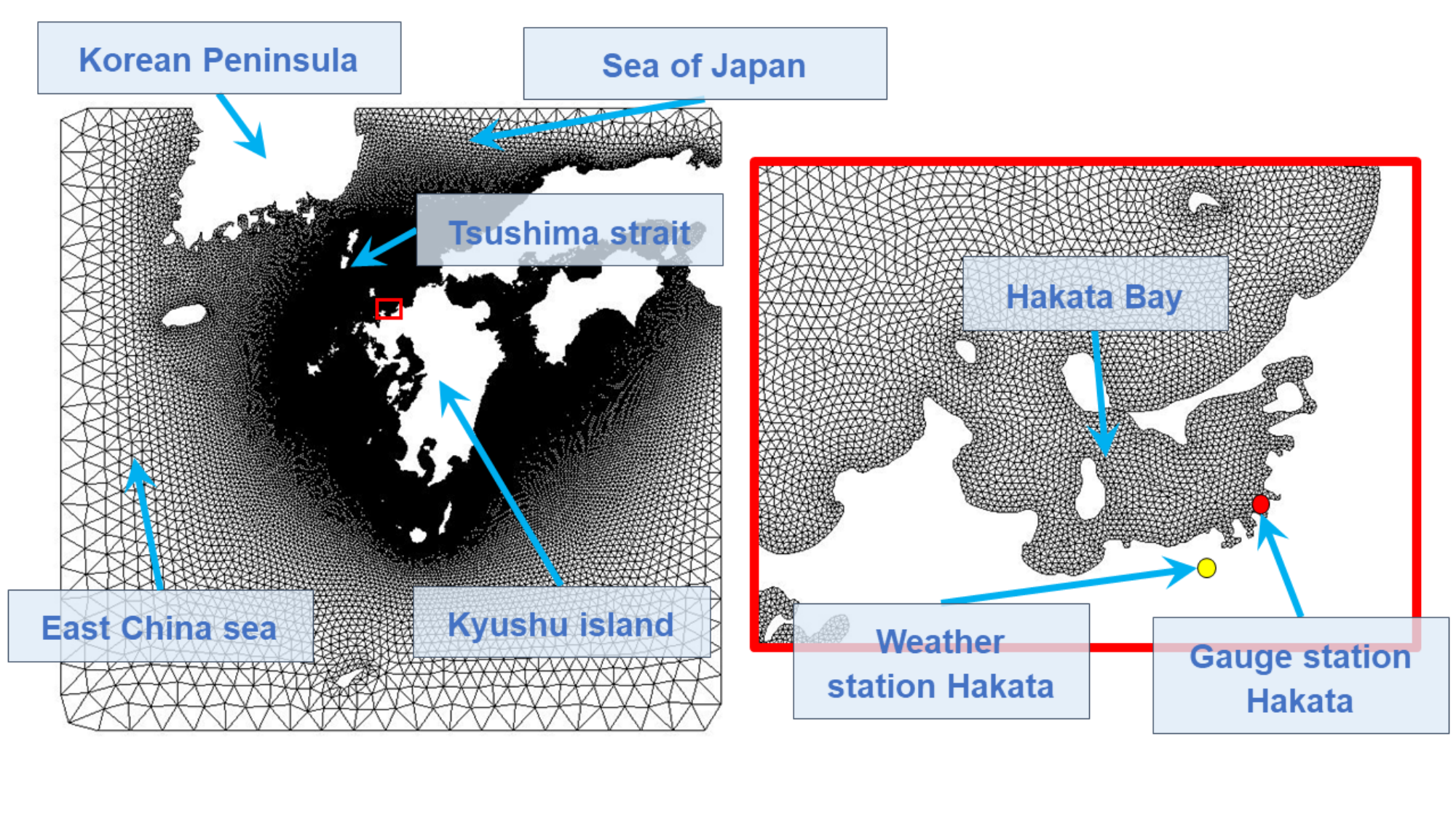}
    \caption{The domain and grids of FVCOM.}
    \label{fig:mesh}
\end{figure}

Figure~\ref{fig:mesh} shows the domains and grids of FVCOM. The coastline was created based on the National Land Numerical Information data from MLIT, and the water depth data were interpolated to each grid point based on the Japan Hydrographic Association's digital seafloor topography data and the Japan Oceanographic Data Center's seafloor topography data. The grid spacing is set to a maximum width of 50~km at the open boundary and a minimum width of 300~m along the coastline of Kyushu Island. The vertical layer division number in the sigma coordinate system is three layers. For the sea surface boundary conditions, the ocean surface wind and atmospheric pressure from the grid point value data of the numerical weather prediction (Meso-Scale Model, MSM) are provided as forces external to the sea surface. Three cases were considered for each typhoon.
\begin{enumerate}
\item Providing wind and atmospheric pressure simultaneously.
\item Providing atmospheric pressure only.
\item Providing wind only.
\end{enumerate}
Additionally, T9918 was excluded from the calculations because MSM data were not available for this typhoon.

\subsection{Result}
\subsubsection{Validation at Hakata Bay tidal gauge station}

\begin{figure}
    \centering
    \includegraphics[width=\linewidth]{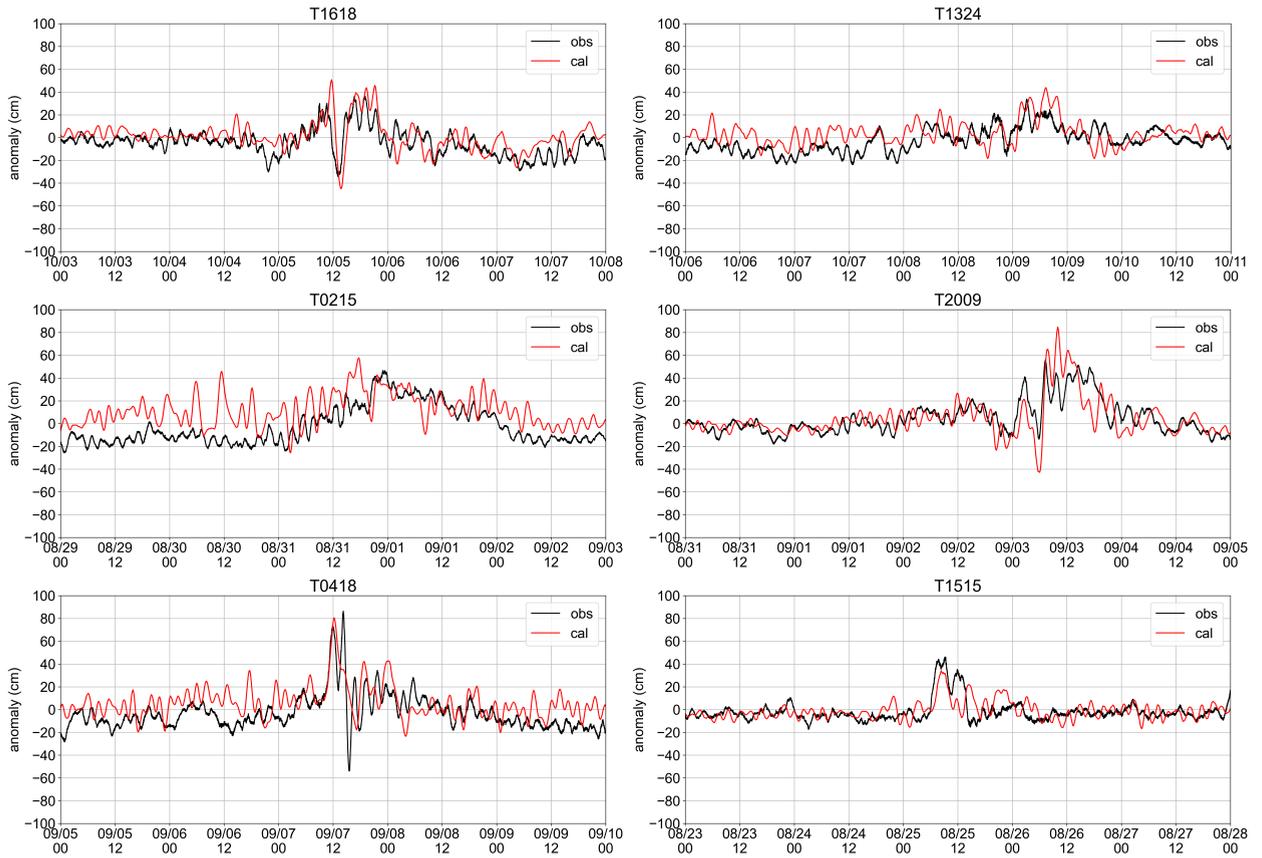}
    \caption{Comparison of simulation results and observations in Hakata Bay}
    \label{fig:compare_obc_cal}
\end{figure}

To validate the storm surge simulation model, the simulated values for Hakata Bay were compared with the storm surge anomalies of the observed values.
Figure~\ref{fig:compare_obc_cal} presents a comparison of the simulation results and observations for Hakata Bay. While the simulation for T0215 exhibits an overestimation of the precursory oscillations before the typhoon's approach, the model accurately represents harbor oscillations in Hakata Bay for other typhoons. The simulation results for T1618 and T2009 capture the characteristics of the first peak around the time the typhoon makes landfall on the Korean Peninsula, followed by a rapid decrease and a subsequent sharp increase to achieve the second peak. Additionally, T1324, which passes close to the northern coast of Kyushu Island, is slightly overestimated at approximately 10:00 on October 9, but generally captures the observed features. For the directly passing overhead typhoons T0418 and T1515, the second peaks are underestimated, but the model reproduces the rapid onset of the first deviation reasonably well.

Considering the differences in validation for each typhoon mentioned above, we consider the characteristics of storm surge development for each typhoon’s trajectory pattern using representative examples. Specifically, for the northeastward trajectory pattern, T1618 and T1324 will be used for northeastward trajectory patterns. For the northward trajectory pattern, T2009 will be considered. Finally, for the direct passing overhead pattern, T1515 will serve as a representative case.
\clearpage

\subsubsection{Contribution of atmospheric pressure and wind to the storm surge}

\begin{figure}
    \centering
    \includegraphics[width=\linewidth]{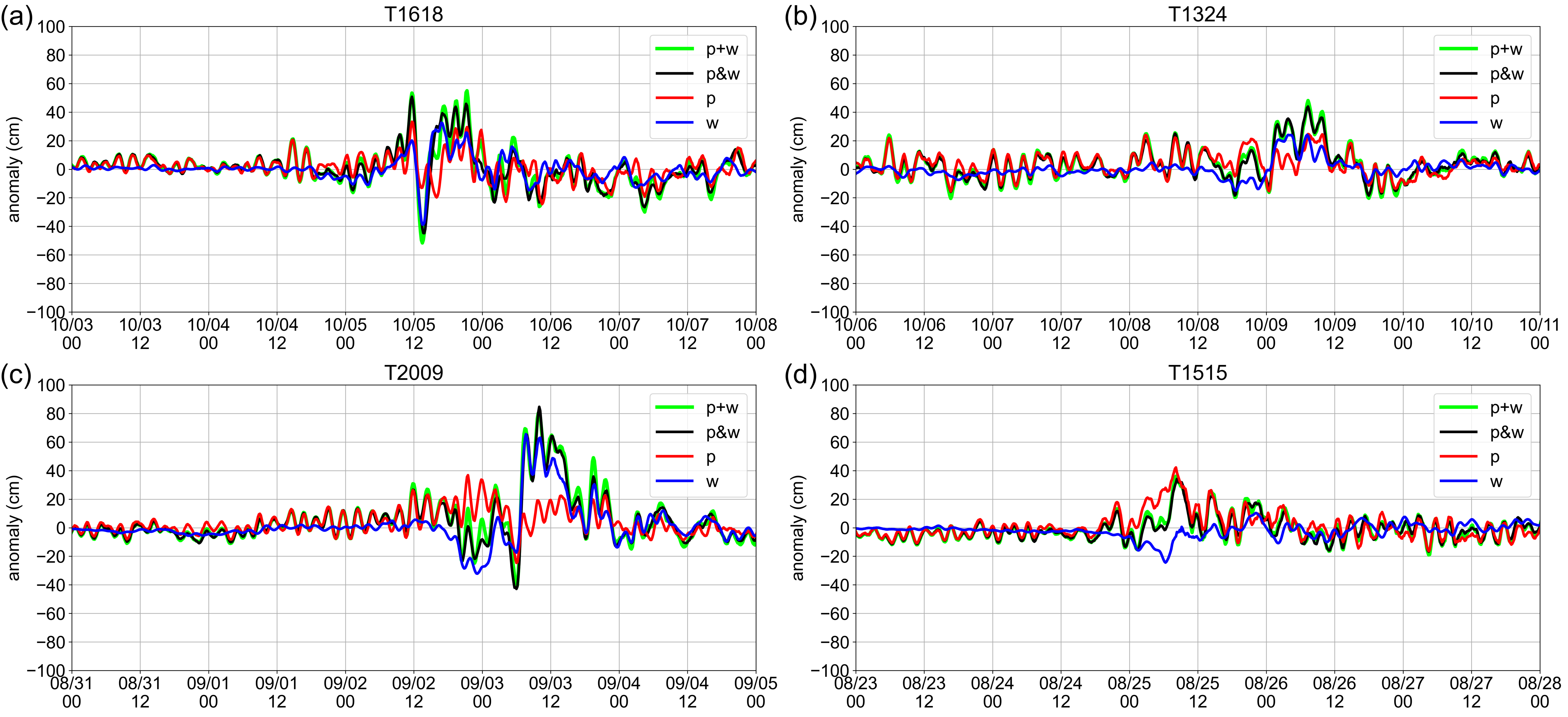}
    \caption{The time series of storm surge anomalies in Hakata Bay under different external forcings applied.}
    \label{fig:cal_hakata_pw}
\end{figure}

We investigated the contributions of wind and atmospheric pressure to storm surges in Hakata Bay. Figure~\ref{fig:cal_hakata_pw} shows the time series of storm surge anomalies in Hakata Bay under different external forcings of T1618, T1324, T2009, and T1515. The black line (p\&w) represents the case where both atmospheric pressure and wind are applied as external forces, the blue line (w) represents the case where only wind is applied, and the red line (p) represents the case where only atmospheric pressure is applied, and the green line (p+w) represents the sum of p and w.
For all typhoons, the case (p+w) is almost equal to the case in which both atmospheric pressure and wind are applied (p\&w). Therefore, in Hakata Bay, storm surge anomalies induced by atmospheric pressure and wind appear to influence each other minimally, allowing for the independent consideration of storm surge development mechanisms for each external force.
\clearpage

\section{Discussion of storm surge development mechanisms}\label{Discussion of storm surge development mechanisms}
\subsection{Natural oscillation in Tsushima Strait}\label{Natural oscillation in Tsushima Strait}

\begin{figure}
    \centering
    \includegraphics[width=0.5\linewidth]{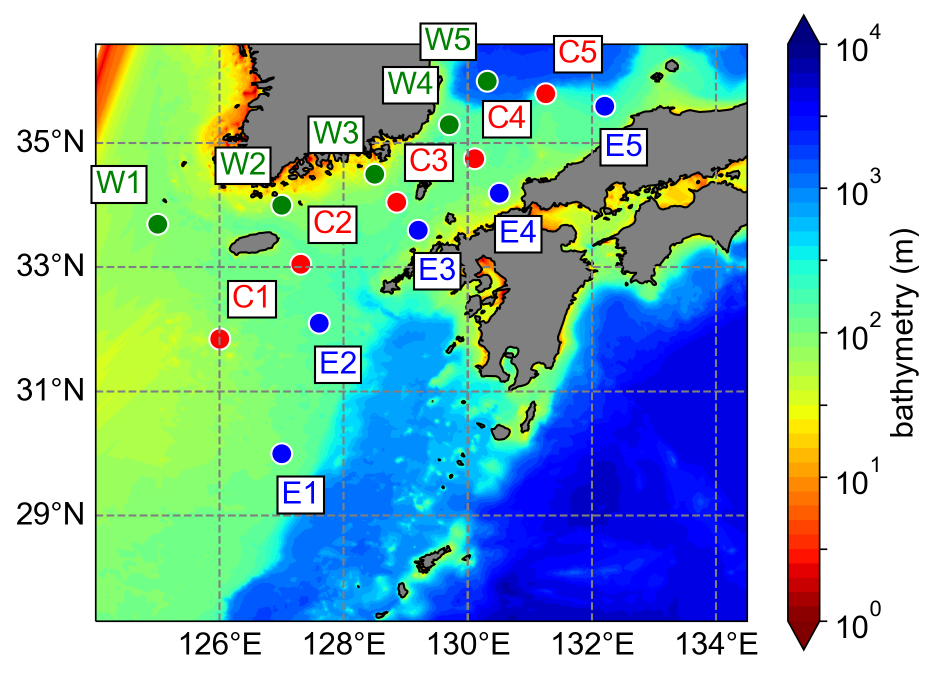}
    \caption{Location of the 15 points}
    \label{fig:stations_bathymetry}
\end{figure}
To investigate the reasons for the occurrence of the 10~hour periodic oscillations observed at the coastal observation points along the Tsushima Strait, spectral analysis was conducted to assess the calculated values at both Hakata Bay and the 15 points shown in Figure~\ref{fig:stations_bathymetry}.
\begin{figure}
    \centering
    \includegraphics[width=0.5\linewidth]{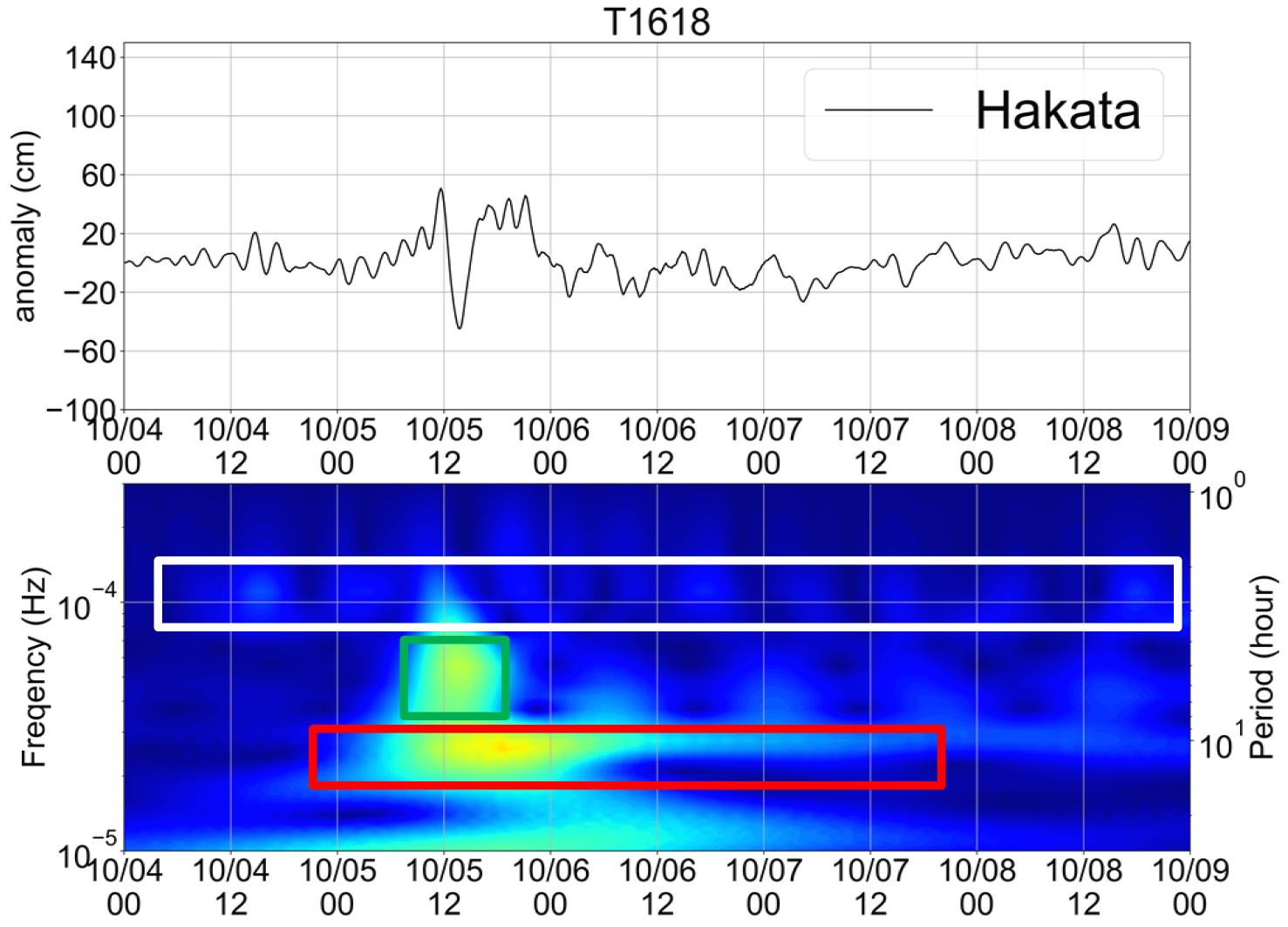}
    \caption{The time series and wavelet transform of the calculated storm surge anomaly in Hakata Bay by T1618.}
    \label{fig:cal_wavelet_hakata_T1618}
\end{figure}
Figure~\ref{fig:cal_wavelet_hakata_T1618} shows the time series and wavelet transform of the values simulated for Hakata Bay for T1618. Similar to the observational results, the simulation results indicate the presence of a 2~hour periodic oscillation in Hakata Bay (white box). Additionally, a 7~hour periodic wave (green box) is observed from the time of the first peak until the occurrence of the second peak. Furthermore, the 10~hour periodic wave (red box) is present in the bay at least one day before the typhoon's closest approach, reaching its maximum amplitude at the time of the second peak and gradually decreasing thereafter, lasting for at least two days.

\begin{figure}
    \centering
    \includegraphics[width=0.8\linewidth]{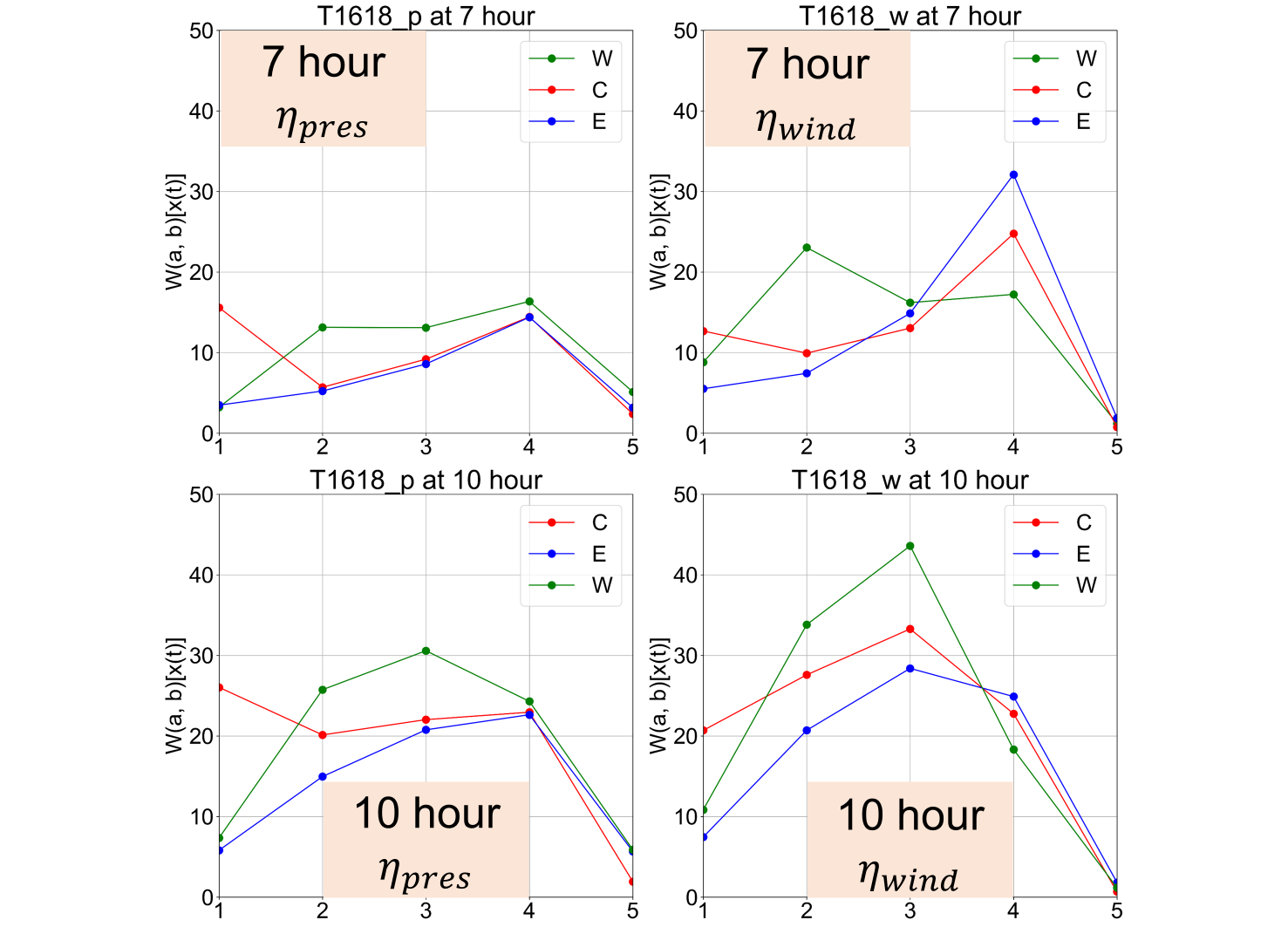}
    \caption{The maximum values of the wavelet transform at the 15 observation points by T1618 when atmospheric pressure and wind were applied as external forces.}
    \label{fig:cal_wavelet_all_points_T1618}
\end{figure}

The maximum values of the wavelet transform of the storm surge anomalies using T1618, considering both atmospheric pressure and wind as external forces at the 15 observation points shown in Figure~\ref{fig:stations_bathymetry} are presented in Figure~\ref{fig:cal_wavelet_all_points_T1618}. Note that all 15 points are located outside Hakata Bay, and so the 2~hour periodic tidal oscillations observed in Hakata Bay (as seen in Figure~\ref{fig:cal_wavelet_hakata_T1618}: white box) are not present at these points.
The 7~hour periodic oscillations are most prominently detected at points W4, C4, and E4 when wind is applied as an external force. Therefore, the 7~hour periodic oscillations are considered to be primarily driven by wind-induced forced waves.
Moreover, for the 10~hour periodic waves, the largest amplitudes are observed when the wind is the external force, particularly at points W3, C3, and E3. In contrast, no significant 10~hour periodic oscillations are observed at points W1, E1, W5, C5, and E5. This suggests that W3, C3, and E3, located near the central part of the Tsushima Strait, correspond to the antinodes of the 10~hour periodic oscillations, while the oscillations decrease in magnitude as one moves closer to the Sea of Japan or the East China Sea, forming an oscillatory pattern with nodes along the boundary of the strait.

The formation of nodes near the boundary with the Sea of Japan or the East China Sea is closely related to water depth. As shown in Figure~\ref{fig:stations_bathymetry}, there is a significant change in water depth at the boundary between the Sea of Japan and the Tsushima Strait. The water depth at points W3, C3, and E3 is approximately 100~m, whereas at W5 and C5 it is 2000~m, and at E5 it is 300~m, thus indicating significantly deeper depths compared to the strait. In cases where the water depth exhibits a discontinuity, the amplitude of the oscillations near the discontinuity point is known to exponentially decrease. For instance, the waveform of edge waves in the case of a discontinuity in the water depth is given by the following equation:
\begin{align}
    \eta_{1}&=a\cos(\mu_{1}x)\cos(ky+\sigma t) \\
    \eta_{2}&=a\cos(\mu_{1}l)e^{-\mu_{2}(x-l)}\cos(ky+\sigma t)
    \label{Edge wave} \\
    \mu_{1}&=\sqrt{\frac{\sigma^2}{gh_{1}} - k^2} \\
    \mu_{2}&=\sqrt{k^2 - \frac{\sigma^2}{gh_{2}}}
\end{align}
Here, we consider the coastline as the \(y\)-axis and the sea as \(x>0\), where \(h_1\) and \(l\) represent the depth and width of the continental shelf, respectively, and \(h_2\) is the depth of the outer region, where the depths are constant with \(h_1<h_2\). In addition, \(k\) and \(\sigma\) are the wavenumber and angular frequency, respectively, and \(a\) is a constant.

According to Equation (\ref{Edge wave}), the amplitude in the region with a greater depth \(x>l\) is significantly smaller than that in the region with a shallower depth \(x<l\). This phenomenon is attributed to the nature of long waves. The group velocity of long waves, expressed as \(C=\sqrt{gh}\), is lower in shallow regions and higher in deep regions. Consequently, when long waves generated in shallow areas propagate into deeper regions, the group velocity increases rapidly. This abrupt change in the group velocity disturbs wave propagation, causing wave reflection at the point of depth change. The reflected and incident waves interfere with the creation of a standing wave, forming a fluid oscillation system in which specific periodic oscillations dominate, which is similar to the resonance in a cup of tea. Therefore, points with very deep depths, such as W5, C5, and E5, do not oscillate but become nodes, and the 10~hour periodic oscillation is expected to occur only in shallow regions.

It has been observed that the 10~hour periodic oscillation exhibits a spatial scale with nodes at W3, C3, and E3 and antinodes at W1, E1, W5, C5, and E5, as shown in Figure~\ref{fig:cal_wavelet_hakata_T1618}. Furthermore, these oscillations have been consistently present since the closest approach of typhoons. The spacing between C3 and C5 is approximately 293~km, whereas that between W3 and E3 is approximately 200~km. The oscillation mode in the direction across the Tsushima Strait from the northern coast of Kyushu Island to the Korean Peninsula is characterized by \(n=0\) because there is no phase difference between W3 and E3. Additionally, in the longitudinal direction of the Tsushima Strait, the oscillation mode is \(m=1\) as antinodes are observed at W3, C3, and E3, and nodes are observed at W1, E1, W5, C5, and E5. Therefore, by considering a rectangular with uniform depth \(h\), length \(l\), and width \(b\), where \(l=293\times2\)~km, \(b=200\)~km, \(m=1\), \(n=0\) and \(h=100\)~m, the following period of free oscillation 
\begin{equation}
    T_{m, n} = \frac{2}{\sqrt{gh}} \left\{\left(\frac{m}{l}\right)^2+\left(\frac{n}{b}\right)^2\right\}^{-1/2}\label{natual ocs}
\end{equation}
is calculated to be \(T_{1, 0}=10.4\)~hours. This corresponds to the observed and simulated periods. In other words, the 10~hour periodic oscillation represents a wave oscillating throughout the entire Tsushima Strait, behaving as a one-dimensional free oscillation with the boundary between the Sea of Japan and the Tsushima Strait acting as its node.

\subsection{Northeastward-moving types}

In this section, we examine the mechanism of storm surge development on the northern coast of Kyushu Island caused by northeast-moving typhoons, based on the spatial distribution of storm surges and external forces. The analysis of observational data from Hakata Bay revealed significant variations depending on whether a northeast-moving typhoon passed closer to the southern coast of the Korean Peninsula or the northern coast of Kyushu. Therefore, we used T1618 as an example that passed closer to the southern coast of the Korean Peninsula, and T1324 as an example that passed closer to the northern coast of Kyushu Island.

\subsubsection{T1618}

\begin{figure}
    \centering
    \includegraphics[width=0.6\linewidth]{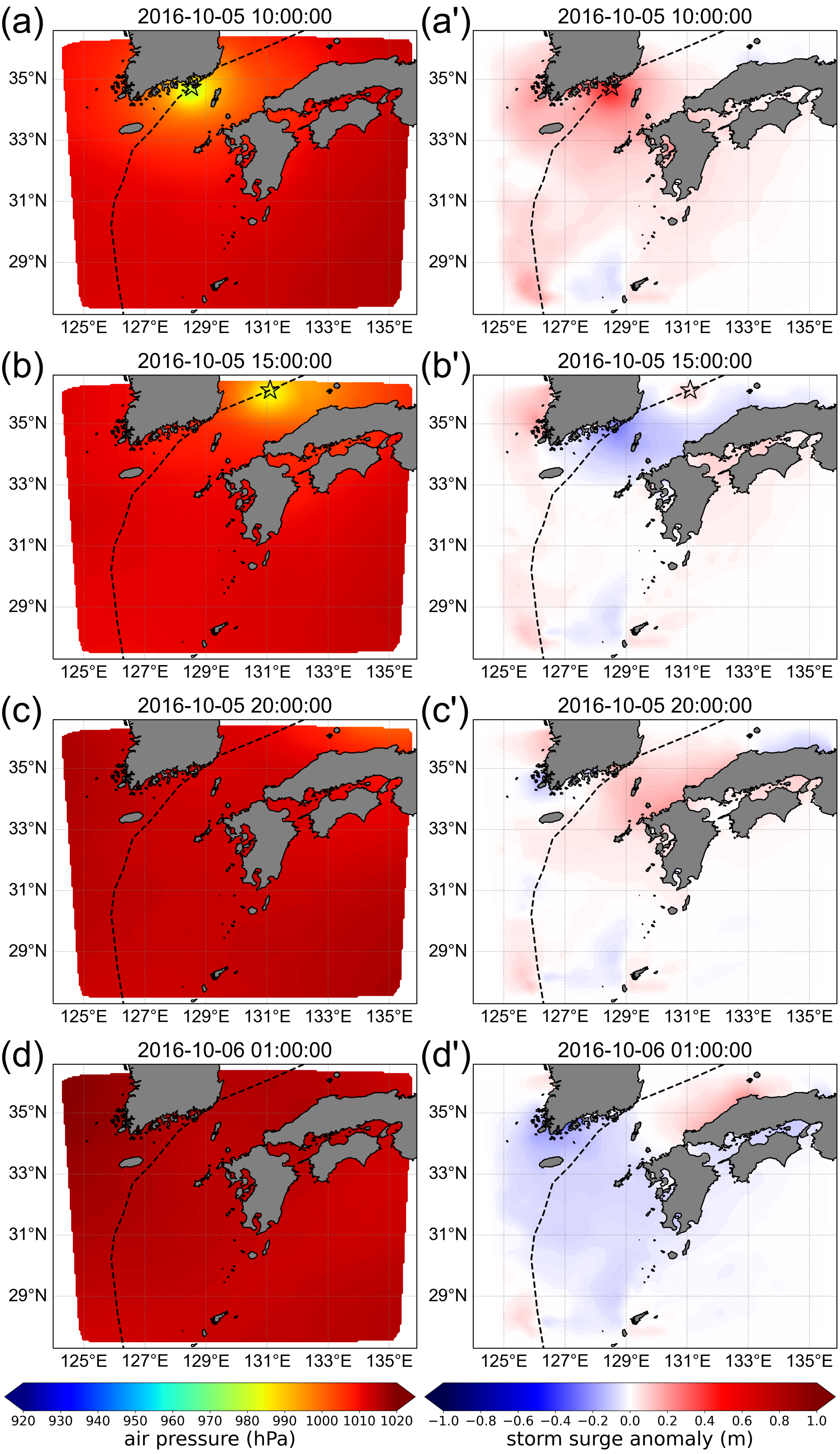}
    \caption{Spatial distribution of atmospheric pressure (left) and storm surge induced by atmospheric pressure (right) in the case of T1618. The dashed line represents the typhoon track, and the star indicates the center position of the typhoon.}
    \label{fig:dist_T1618_p}
\end{figure}
For typhoons following a path that passes near the southern coast of the Korean Peninsula (T0314, T0415, T1618, T1825, T2211), the storm surge reaches its first peak at the time of the typhoon's closest approach, and then rapidly decreases to a minimum. Subsequently, after the typhoon moves away from the northern coast of Kyushu Island, the storm surge rises again, reaching a second peak. In the current section, we aim to identify the mechanism underlying this storm surge behavior.

First, considering only atmospheric pressure as the external force, we examine the storm surge development mechanism along the northern coast of Kyushu Island based on the spatial distribution of the sea-level anomaly induced by atmospheric pressure. Figure~\ref{fig:dist_T1618_p} illustrates the spatial distribution of the storm surge anomaly when the atmospheric pressure from Typhoon T1618 is applied. On October 5, 2016, at 10:00, the center of the typhoon was located near the southern coast of the Korean Peninsula, and atmospheric pressure was distributed concentrically around the typhoon's center (Figure~\ref{fig:dist_T1618_p}a). Corresponding to this concentric low-pressure distribution, the storm surge anomaly rose concentrically across the entire Tsushima Strait because of the inverted barometer effect (Figure~\ref{fig:dist_T1618_p}a'). At this time, the northern coast of Kyushu Island experienced a positive sea-level anomaly of 30~cm. However, in Hakata Bay, 2 hours’ oscillation occurred at sea level, resulting in the first peak occurring at around noon on October 5, 2016, approximately 2 hours from the time of the typhoon's closest approach (Figure~\ref{fig:cal_hakata_pw}a: p).

On October 5, 2016, at 15:00, the typhoon advanced to a northeast direction, entering the Sea of Japan. At this time, the atmospheric pressure over the Tsushima Strait increased to approximately 1,013~hPa (Figure~\ref{fig:dist_T1618_p}b). Consequently, the lifted seawater in the Tsushima Strait was released, and the potential energy was liberated, initiating the 10~hour periodic natural oscillation of the Tsushima Strait discussed in Subsection \ref{Natural oscillation in Tsushima Strait}. Therefore, from 10:00 on October 5, 2016, to half a period later at 15:00 on the same day, a negative storm surge anomaly occurred in the Tsushima Strait (Figure~\ref{fig:dist_T1618_p}b'). 
Furthermore, half a period later, at 20:00 on October 5, 2016, due to the back-and-forth motion of the Tsushima Strait's natural oscillation, a positive storm surge anomaly developed across the entire Tsushima Strait (Figure~\ref{fig:dist_T1618_p}c'). 
Subsequently, at 01:00 on October 6, 2016, a negative anomaly was observed in Tsushima Strait, whereas a positive anomaly occurred along the Sea of Japan coast.
The reason for the positive anomaly along the coast of the Japan Sea is that, over time, after the initiation of natural oscillations in the Tsushima Strait, the Coriolis force begins to take effect. By 20:00 on October 5, 2016, the positive anomaly that had occurred along the northern coast of Kyushu Island propagated towards the Japan Sea as shelf waves or Kelvin waves.
In general, the timescale at which the Coriolis force becomes effective is given by 
\begin{align}
    T=\frac{24}{2\sin(\phi)}
    \label{Coriolis time scale}
\end{align}
Here, \(\phi\) represents the latitude and substituting \(\phi=30^{\circ}~N\), \(T\) becomes 24 hours.

\begin{figure}
    \centering
    \includegraphics[width=\linewidth]{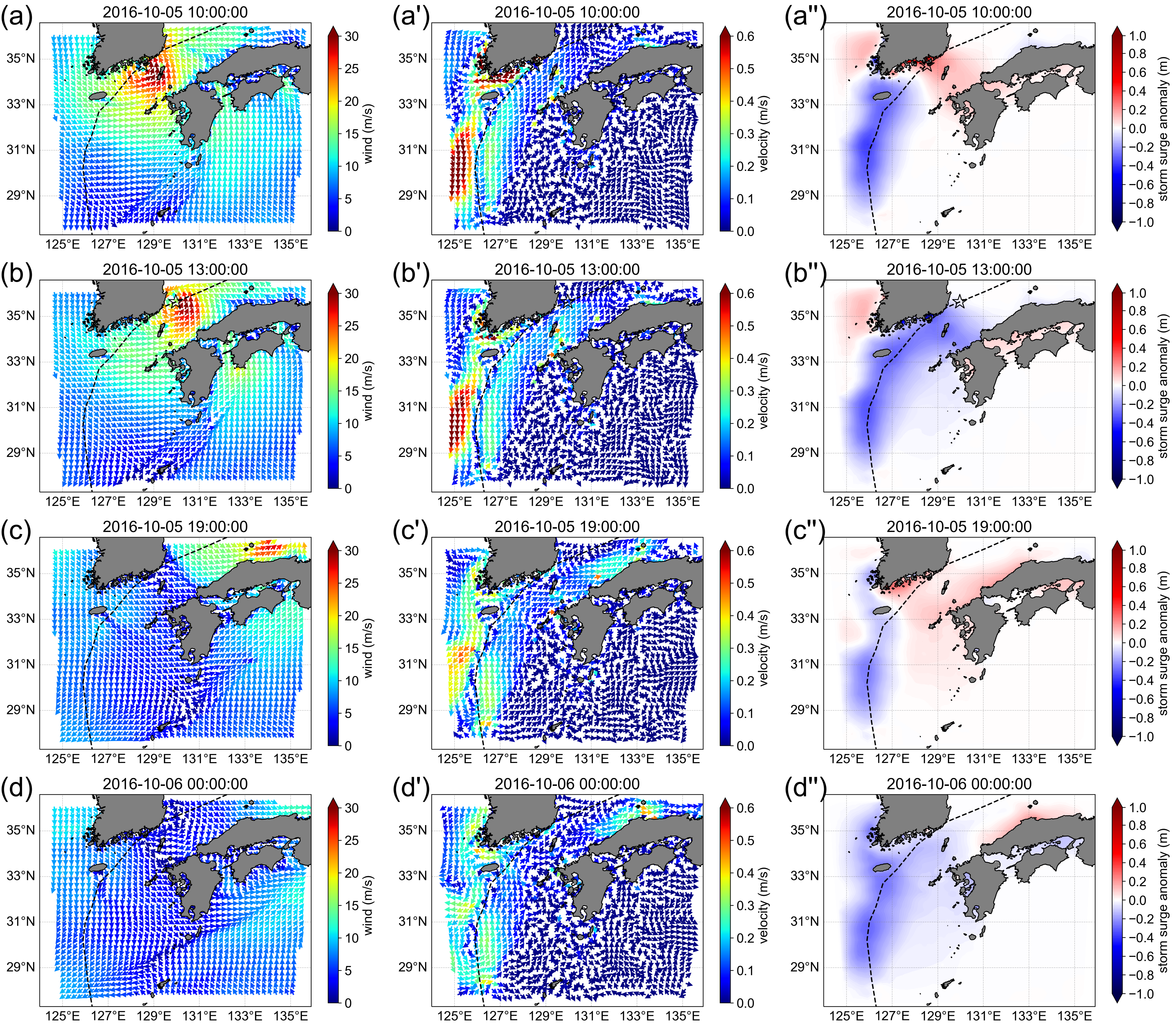}
    \caption{Spatial distribution of storm surge anomalies and vertically averaged flow velocity induced by the wind in the case of T1618 (Left: Wind, Middle: Vertically averaged flow velocity, Right: storm surge anomalies). The dashed line represents the typhoon track, and the star indicates the center position of the typhoon.}
    \label{fig:dist_T1618_w}
\end{figure}

Next, by providing only wind as an external force, the spatial distribution of storm surge anomalies generated by wind was used. Figure~\ref{fig:dist_T1618_w} illustrates the spatial distribution of the storm surge anomaly and vertically averaged flow velocity when T1618's wind is applied. At 10:00 on October 5, 2016, the typhoon was located off the southern coast of the Korean Peninsula, creating a counterclockwise wind field around the typhoon center in the Tsushima Strait (Figure~\ref{fig:dist_T1618_w}a). During this time, positive anomalies occurred in the Tsushima Strait, whereas behind the typhoon's path, Ekman pumping induced negative storm surge anomalies (Figure~\ref{fig:dist_T1618_w}a'').
Ekman pumping is the divergence of seawater associated with wind distribution. When a counterclockwise wind field acts on seawater, it induces counterclockwise flow. In the Northern Hemisphere, owing to the Coriolis effect acting to the right of the flow, after a sufficient amount of time elapses, the seawater flow changes direction away from the center of circulation, leading to the generation of negative storm surge anomalies. This circulation persists even after the typhoon passes, resulting in a prolonged negative anomaly along the path taken by the typhoon.

The cause of the positive storm surge anomaly in the Tsushima Strait is considered. Near Tsushima Island (\(129^{\circ}~E, 34^{\circ}~N\)), as a typhoon approaches, the wind direction changes northeastward and the wind speed increases to 25~m/s or more, leading to a prevailing northeasterly flow in the seawater (Figure~\ref{fig:dist_T1618_w}a'). However, on the eastern side of Tsushima Strait (\(131^{\circ}~E, 35^{\circ}~N\)), northwestward winds create a weak northwestward flow (Figure~\ref{fig:dist_T1618_w}a'). As a result, the vicinity of Tsushima Island becomes a convergence zone for seawater, causing seawater to stagnate in the Tsushima Strait and leading to a positive storm surge anomaly.

On October 5, 2016, at 13:00, as the typhoon moved northeastward, a significantly negative storm surge anomaly occurred throughout the Tsushima Strait, reaching a minimum peak on the northern coast of Kyushu Island (Figure~\ref{fig:dist_T1618_w}b''). Winds on the east side of the Tsushima Strait (\(131^{\circ}~E, 35^{\circ}~N\)) became strong and northward, causing a northeastward flow of seawater (Figure~\ref{fig:dist_T1618_w}b, b'). Consequently, the stagnated seawater in the Tsushima Strait was released, and oscillations began with the release of potential energy, leading to a negative storm surge anomaly in the Tsushima Strait. However, the oscillation-induced negative anomaly triggered by an increase in atmospheric pressure required 5 hours from the release of potential energy to the occurrence of a negative anomaly (Figure~\ref{fig:dist_T1618_p}a', b'). In contrast, the oscillation induced by wind resulted in a negative anomaly that occurred 3 hours after the release of potential energy. This time difference was attributed to the rapid change in flow direction in the Tsushima Strait, which allowed for the swift release of potential energy. Therefore, when wind was the external force, a 7~hour periodic oscillation, slightly shorter than the intrinsic oscillation of the Tsushima Strait, occurred from the first peak until the second peak. After the second peak, the influence of topography prolonged the period, leading to the dominance of the Tsushima Strait's 10~hour intrinsic oscillation.

On October 5, 2016, at 19:00, as the typhoon moved into the Sea of Japan, a significant positive storm surge anomaly occurred not only in the Tsushima Strait but also along the coastal areas of the Sea of Japan, reaching its second peak on the northern coast of Kyushu Island (Figure~\ref{fig:dist_T1618_w}c''). The positive anomaly generated in the Tsushima Strait is believed to be the result of the oscillation rebound induced by the release of potential energy. Meanwhile, southeastward winds prevail from the northern coast of Kyushu Island to the coastal areas of the Sea of Japan (Figure~\ref{fig:dist_T1618_w}c). As these southeastward winds blow, the direction of the seawater flow shifts from northeastward to east-northeastward (Figure~\ref{fig:dist_T1618_w}c'). Consequently, the pushing of seawater toward the northern coast of Kyushu Island and the coastal areas of the Sea of Japan, accompanied by Ekman transport, leads to the occurrence of positive storm surge anomalies in both the Tsushima Strait and the Sea of Japan.

\begin{figure}
    \centering
    \includegraphics[width=0.75\linewidth]{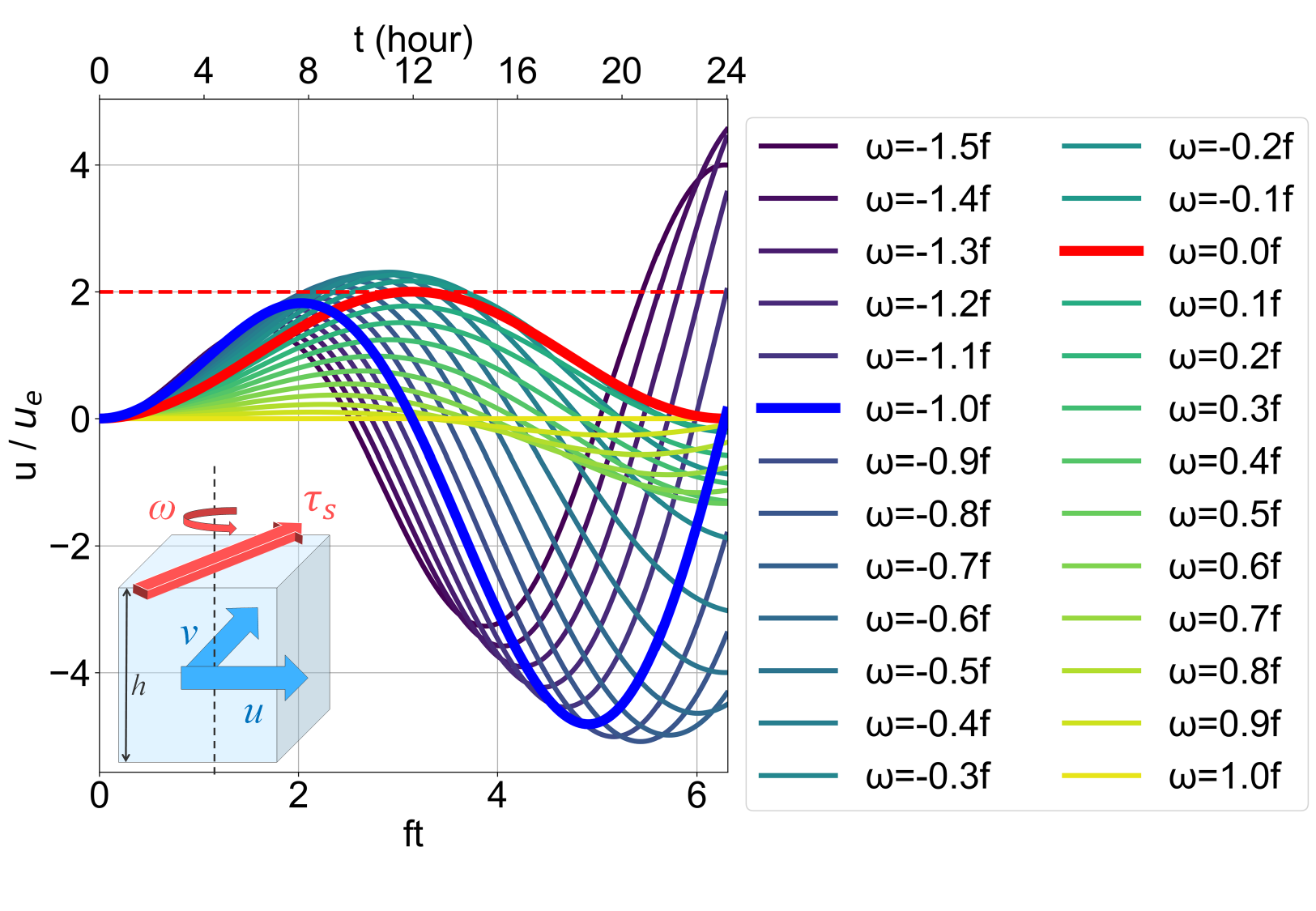}
    \caption{Ekman transport by the wind the direction of which rotate}
    \label{fig:ekman_u}
\end{figure}

Generally, Ekman transport is a phenomenon in which seawater is transported to the right of the wind direction due to prolonged winds blowing in the same direction. Therefore, we consider Ekman transport in cases where the direction of wind rotation changes as a typhoon passes, as observed in this scenario. Consider wind acting on a water column of height \(h\) and a unit area rotating at a constant speed (Figure~\ref{fig:ekman_u}: bottom left). If we denote the rotational speed of the wind as \(\omega\) and the wind speed as \(V\), the vertically averaged momentum equation is as follows:
\begin{align}
    \frac{\partial u}{\partial t}  - fv = \frac{\tau_{sx}}{\rho h} = \frac{\rho_a C_s V^{2}}{\rho h}\cos{\omega t} \label{u_av_eq} \\
    \frac{\partial v}{\partial t}  + fu = \frac{\tau_{sy}}{\rho h} = \frac{\rho_a C_s V^{2}}{\rho h}\sin{\omega t} \label{v_av_eq}
\end{align}
The initial conditions at \(t = 0\) are as follows:
\begin{align}
    u = v = 0 \label{IC}
\end{align}
By solving equations (\ref{u_av_eq}) and (\ref{v_av_eq}) based on the initial conditions (\ref{IC}), the following solutions are obtained:
\begin{align}
    u = \frac{2\rho_a C_s V^{2}}{\rho h (f+\omega)} \cos{\left(\frac{f-\omega}{2}t\right)} \sin{\left(\frac{f+\omega}{2}t\right)} \label{u_av} \\
    v = \frac{-2\rho_a C_s V^{2}}{\rho h (f+\omega)} \sin{\left(\frac{f-\omega}{2}t\right)} \sin{\left(\frac{f+\omega}{2}t\right)} \label{v_av}
\end{align}
Figure~\ref{fig:ekman_u} illustrates Equation (\ref{u_av}) for various rotational speeds. Here, we assume a latitude of \(30^{\circ}~N\) and the initial wind direction is along the \(y\)-axis. The magnitude is normalized by the magnitude of the vertical average flow when steady wind is applied (equation \ref{u_e}).
\begin{align}
    u_{e} = \frac{\rho_a C_s V^{2}}{f \rho h} \label{u_e}
\end{align}

The red line represents the case in which steady wind is applied. In the case of steady wind, the flow velocity gradually increases from the start of the calculation, reaching a peak value of 2.0 at around 12 hours, following which the velocity decreases, reaching zero at 24 hours. 
However, when the wind rotates counterclockwise (\(\omega>0\)), the flow velocity gradually increases from the start of the calculation, but the peak value is smaller than that in the case of steady wind. 
Additionally, when the wind rotates clockwise (\(\omega<0\)), the flow velocity increases gradually from the start of the calculation, reaching a peak faster than in the case of steady wind. For example, in the case \(\omega=-f\), the peak is reached 8 hours after the start, and the peak value is almost the same as that in the case of steady wind. Moreover, the case with the largest peak is \(\omega=-0.3f\) and the peak value is 2.3.

The reason why the Ekman transport is larger and reaches its peak faster in the case of clockwise rotation compared to the steady wind is interpreted as follows. If we consider the Coriolis force bending the flow as a force limiting the transport of seawater, then the flow generated by the wind blowing steadily in one direction is subject to this limitation. In contrast, when the flow follows the wind direction change due to the Coriolis force, the flow is not subject to the restriction imposed by the Coriolis force. Therefore, when a typhoon moves northeastward or northward through the Tsushima Strait, the wind blowing through the strait is expected to rotate clockwise. Consequently, relatively early, the flow is redirected, leading to Ekman transport, and positive anomalies are generated along the northern coast of Kyushu Island and the coast of the Sea of Japan.

At 00:00 on October 6, 2016, weak westward winds developed near Tsushima Island (Figure~\ref{fig:dist_T1618_w}d). This change in wind direction is considered to be caused by the squeeze flow between the Korean Peninsula and Kyushu Island acting as a barrier. The westward wind counteracts the northeasterly flow that was occurring in the Tsushima Strait, causing the flow to disappear in the strait (Figure~\ref{fig:dist_T1618_w}d'). Consequently, the magnitude of the anomaly decreases in the Tsushima Strait (Figure~\ref{fig:dist_T1618_w}d''). 
In contrast, positive anomalies are maintained in places such as Susa and Hamada (\(133^{\circ}~E, 35^{\circ}~N\)). 
This is because the positive anomalies that occur on the northern coast of Kyushu Island and the coast of the Sea of Japan propagate along the northern coast of Kyushu Island as shelf waves or Kelvin waves. Therefore, in places such as Susa and Hamada (\(133^{\circ}~E, 35^{\circ}~N\)), high anomalies are sustained for an extended period.

\subsubsection{T1324}

Next, let us consider the mechanism of storm surge development for Typhoon T1324, which passed close to the northern coast of Kyushu Island. For northeastern-moving typhoons, such as T1324, T1807, and T1917, the storm surge anomalies in Hakata Bay tended to exhibit a single peak without a rapid decrease in anomalies. In the current section, we aim to identify the mechanism underlying this behavior.

Regarding atmospheric pressure, T1324, like T1618, experiences an increase in pressure as the typhoon passes through the Tsushima Strait. This results in the release of position energy, initiating the 10-hour periodic natural oscillations in the strait.
\begin{figure}
    \centering
    \includegraphics[width=\linewidth]{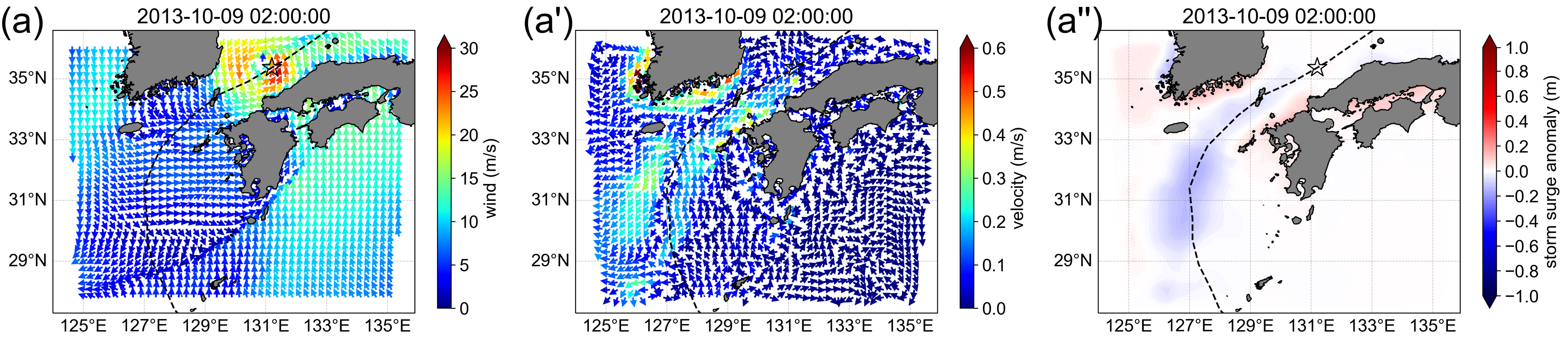}
    \caption{Spatial distribution of storm surge anomalies and vertically averaged flow velocity induced by the wind in the case of T1324 (Left: Wind, Middle: Vertically averaged flow velocity, Right: storm surge anomalies). The dashed line represents the typhoon track, and the star indicates the center position of the typhoon.}
    \label{fig:dist_T1324_w}
\end{figure}

However, what distinguishes T1324 from T1618 is the formation of storm surge anomalies caused by wind. Figure~\ref{fig:dist_T1324_w} illustrates the spatial distribution of storm surge anomalies and the vertical mean flow velocity at 02:00 on October 9, 2013, when wind data from T1324 are applied. In the case of T1324, as the typhoon passes through the Tsushima Strait, negative anomalies occur near Tsushima Island, but positive anomalies develop near the southern coast of the Korean Peninsula and the northern coast of Kyushu Island (Figure~\ref{fig:dist_T1324_w}a''). This is because, as T1324 passes through the central part of the Tsushima Strait, a persistent counterclockwise circulating flow field is established around Tsushima Island, resulting in continued Ekman pumping.
While T1618, passing near the southern coast of the Korean Peninsula, induces a uniform northeastward wind over the Tsushima Strait (Figure~\ref{fig:dist_T1618_w}a), T1324, before approaching the Tsushima Strait, already has southwestward winds persisting near the southern coast of the Korean Peninsula and eastward winds near the northern coast of Kyushu Island (Figure~\ref{fig:dist_T1324_w}a). Consequently, the circulation of flow around Tsushima Island is sustained (Figure~\ref{fig:dist_T1324_w}a'), causing seawater to be pushed toward the southern coast of the Korean Peninsula and the northern coast of Kyushu Island. This results in positive anomalies along the coast of the Tsushima Strait, with negative anomalies occurring around Tsushima Island. 
Subsequently, similar to T1618, the positive anomalies initially present near the northern coast of Kyushu propagate as shelf waves or Kelvin waves along the Sea of Japan, causing a reduction in storm surge anomalies along the northern coast of Kyushu.
\clearpage

\subsection{Northward-moving types}
\begin{figure}
    \centering
    \includegraphics[width=\linewidth]{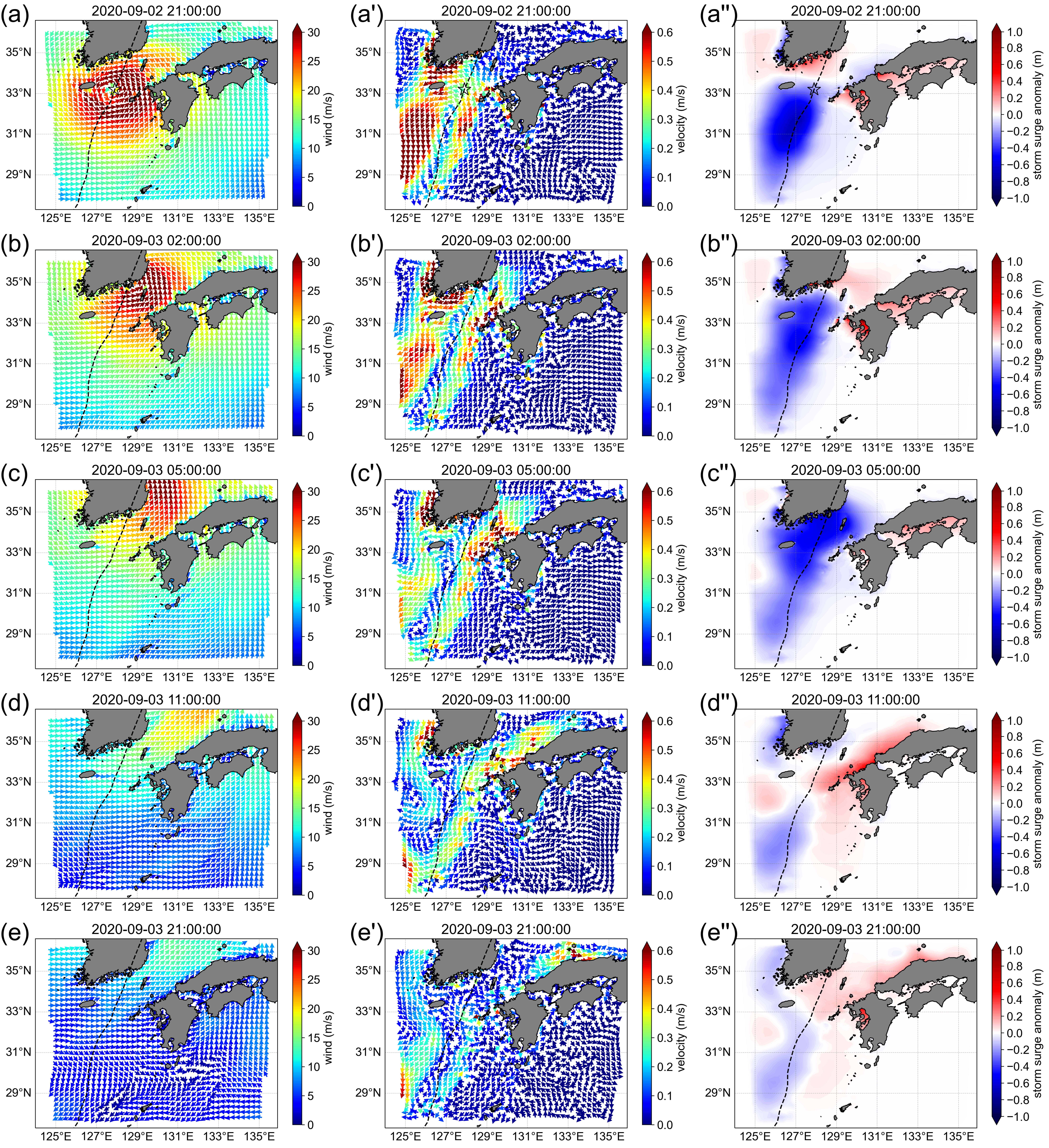}
    \caption{Spatial distribution of storm surge anomalies and vertically averaged flow velocity induced by the wind in the case of T2009 (Left: Wind, Middle: Vertically averaged flow velocity, Right: storm surge anomalies). The dashed line represents the typhoon track, and the star indicates the center position of the typhoon.}
    \label{fig:dist_T2009_w}
\end{figure}

Let us examine the storm surge mechanism of T2009 as a representative of northward-moving typhoons. Storm surges caused by northward-moving typhoons tend to maintain high anomalies for more than 12 hours after the second peak, in contrast with the characteristics of northeastward-moving typhoons.

Similar to T1618, the development mechanism of storm surges caused by atmospheric pressure in T2009 involve a typhoon passing through the Tsushima Strait. As the pressure rises when the typhoon exits the strait, the restraining force that lifts the seawater in the Tsushima Strait is released, initiating the 10~hour periodic natural oscillations in the strait.
However, the mechanism of storm surge development owing to wind is different from that of T1618. Figure~\ref{fig:dist_T2009_w} illustrates the spatial distribution of storm surge anomalies and the vertical mean flow velocity when wind data from T2009 are applied. Immediately before the typhoon enters the Tsushima Strait at 21:00 on September 2, 2020, negative anomalies occur along the northern coast of Kyushu Island, whereas positive anomalies develop near the southern coast of the Korean Peninsula (Figure~\ref{fig:dist_T2009_w}a''). At this time, a prevailing northwestward wind dominates the Tsushima Strait (Figure~\ref{fig:dist_T2009_w}a), causing the flow to face northwest or north (Figure~\ref{fig:dist_T2009_w}a'). The difference in anomalies between the northern coast of Kyushu Island and the southern coast of the Korean Peninsula is caused by the wind setup effect, resulting in positive and negative anomalies, respectively.

When the typhoon made landfall on the Korean Peninsula at 02:00 on September 3, 2020, negative anomalies occurred near the southern coast of the Korean Peninsula, whereas positive anomalies developed along the northern coast of Kyushu (Figure~\ref{fig:dist_T2009_w}b''). At this time, the wind over the Tsushima Strait changed from northwestward to a very strong northeastward direction because the typhoon moved northward (Figure~\ref{fig:dist_T2009_w}b). With this change in wind direction, the flow shifts northeastward, and the flow velocity increases (Figure~\ref{fig:dist_T2009_w}b'). Resultantly, the release of position energy occurs, leading to negative anomalies near the southern coast of the Korean Peninsula and positive anomalies near the northern coast of Kyushu Island, induced by Ekman transport due to rotating winds.
When the typhoon moves northward at 05:00 on September 3, 2020, the Tsushima Strait continues to experience northeastward winds (Figure~\ref{fig:dist_T2009_w}c) and negative anomalies persist around the southern coast of the Korean Peninsula in the strait (Figure~\ref{fig:dist_T2009_w}c''). This is attributed to the ongoing release of position energy from 02:00. The northeastward winds lead to a transition from a strong westward flow to a weaker northeastward flow along the southern coast of the Korean Peninsula, resulting in negative anomalies near the southern coast of the Korean Peninsula (Figure~\ref{fig:dist_T2009_w}c').
By 11:00 on September 3, 2020, positive anomalies had occurred on the northern coast of Kyushu Island and in the coastal areas of the Sea of Japan, whereas negative anomalies developed near the southern coast of the Korean Peninsula (Figure~\ref{fig:dist_T2009_w}d''). Hakata Bay experiences its second peak of storm surge anomalies (Figure~\ref{fig:cal_hakata_pw}c). This distribution of storm surge anomalies is attributed to Ekman transport caused by northeastward winds (Figure~\ref{fig:dist_T2009_w}d). Along the northern coast of Kyushu Island and the coastal areas of the Sea of Japan, a strong northeastward flow is observed, pushing seawater onto the land (Figure~\ref{fig:dist_T2009_w}d').
Until approximately 21:00 on September 3, positive storm surge anomalies persist along the northern coast of Kyushu (Figure~\ref{fig:dist_T2009_w}e''). Unlike northeastward-moving typhoons such as T1618, where the southwestward wind after passage counteracts the northeastward flow due to the typhoon, T2009 continues to have northeastward winds over the Tsushima Strait even after passage. This sustains northeastward wind-induced Ekman transport, resulting in positive anomalies that last for more than 10 hours along the northern coast of Kyushu Island. The slight difference in typhoon paths plays a crucial role in determining whether highly-positive anomalies are sustained after the second peak, enhancing the possibility of overlapping with high tide.
\clearpage

\subsection{Directly passing overhead types}

\begin{figure}
    \centering
    \includegraphics[width=0.5\linewidth]{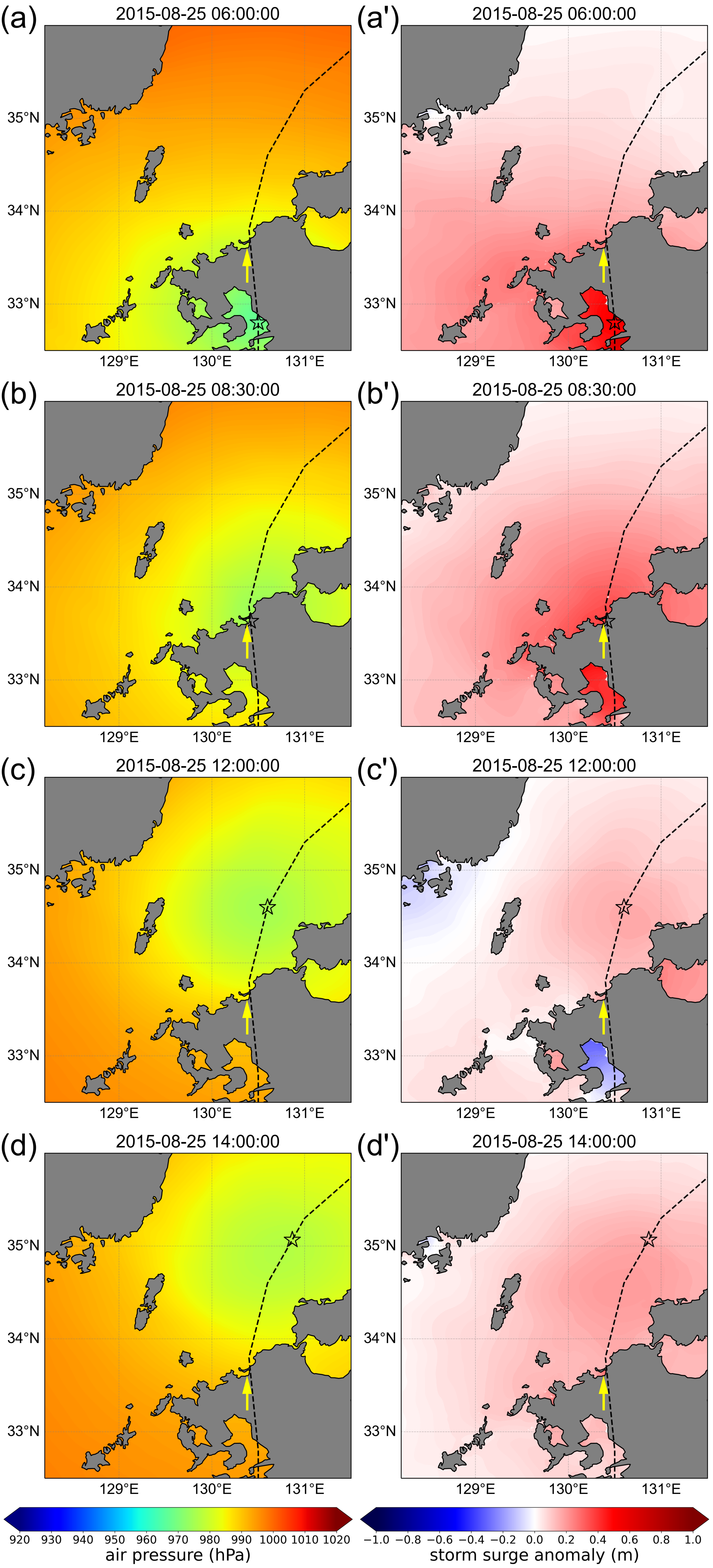}
    \caption{Spatial distribution of atmospheric pressure (left) and storm surge induced by atmospheric pressure (right) in the case of T1515. The dashed line represents the typhoon track, and the star indicates the center position of the typhoon.}
    \label{fig:dist_T1515_p}
\end{figure}

Finally, we consider the storm surge mechanism caused by T1515, which represents the case of a storm passing directly through the overhead. Storm surges induced by such storms exhibit the characteristic of achieving one or two peaks in surge anomalies within a short period during the closest approach to a typhoon. In this section, we identify the mechanisms underlying the aforementioned characteristics.

Figure~\ref{fig:dist_T1515_p} illustrates the spatial distribution of the storm surge anomaly when only atmospheric pressure effects are considered for Typhoon T1515. The map is magnified on the northern coast of Kyushu Island, with the location of Hakata Bay indicated by the yellow arrow. At 6:00 on August 25, 2015, the typhoon was positioned off the west coast of Kyushu Island, leading to a positive surge anomaly centered around the typhoon owing to the inverted barometer effect (Figure~\ref{fig:dist_T1515_p}a, a'). Hakata Bay experienced a surge anomaly of 20~cm (Figure~\ref{fig:cal_hakata_pw}d: p); however, the rapid surge observed in the actual measurements over a short period was not reproduced by atmospheric pressure alone.
At 8:30 on August 25, 2015, the typhoon passed over Hakata Bay. During this passage, the surge anomaly exhibited a distribution of positive anomalies centered around the typhoon due to the inverted barometer effect (Figure~\ref{fig:dist_T1515_p}b'), and Hakata Bay experienced the first peak of the surge anomaly (Figure~\ref{fig:cal_hakata_pw}d: p).
Subsequently, at 12:00 on August 25, 2015, the atmospheric pressure increased as the typhoon moved northward, causing a decrease in surge anomalies. However, in Hakata Bay, an oscillation occurred, resulting in a larger positive anomaly compared to the surrounding areas, leading to the occurrence of the second peak (Figure~\ref{fig:dist_T1515_p}c, c'). At 14:00 on August 25, 2015, Hakata Bay experienced another peak owing to bay oscillation (Figure~\ref{fig:dist_T1515_p}d').

\begin{figure}
    \centering
    \includegraphics[width=\linewidth]{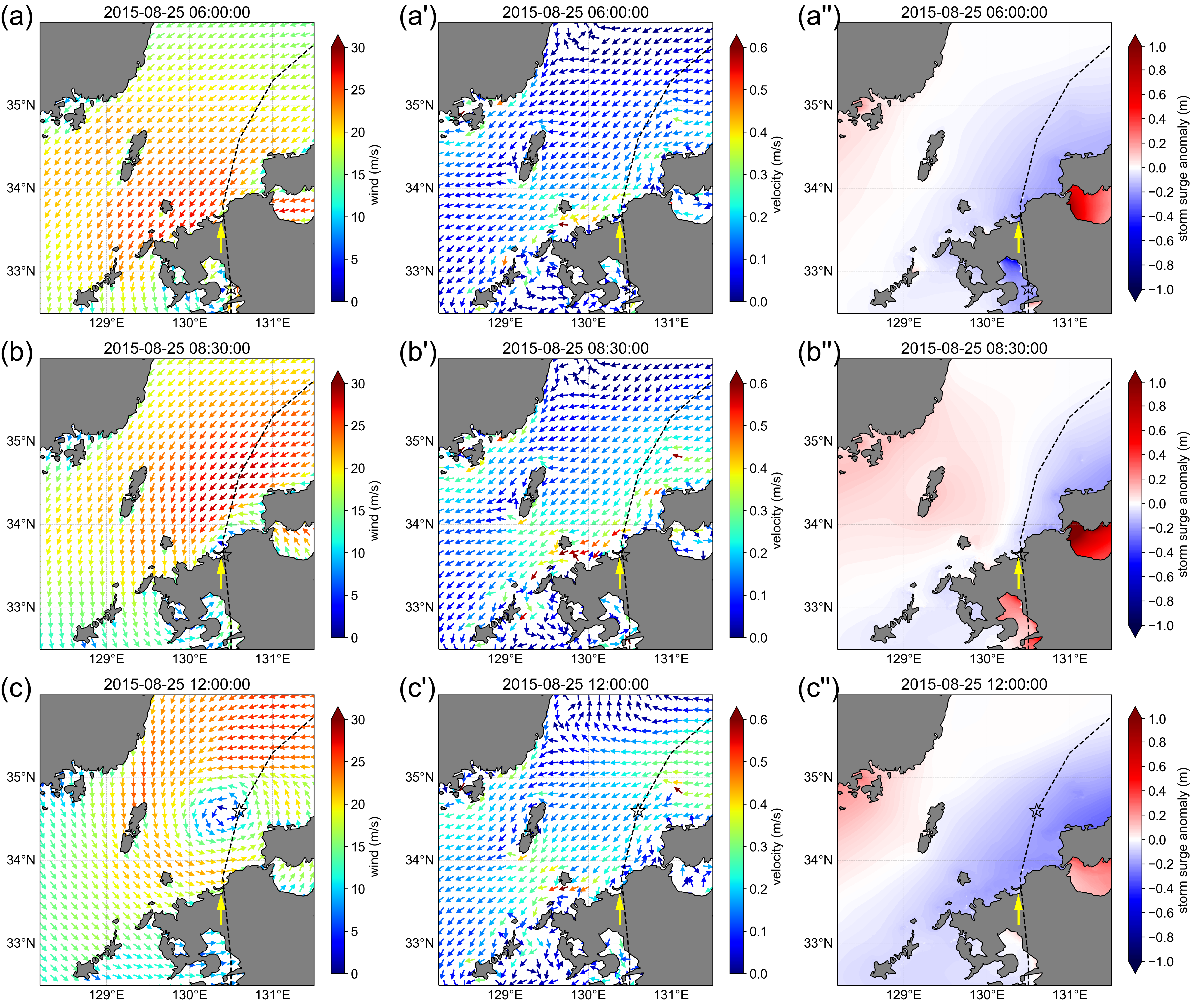}
    \caption{Spatial distribution of storm surge anomalies and vertically averaged flow velocity induced by the wind in the case of T1515 (Left: Wind, Middle: Vertically averaged flow velocity, Right: storm surge anomalies). The dashed line represents the typhoon track, and the star indicates the center position of the typhoon.}
    \label{fig:dist_T1515_w}
\end{figure}

Continuing with the consideration of wind effects, Figure~\ref{fig:dist_T1515_w} illustrates the spatial distribution of vertical average flow velocity and storm surge anomaly when only wind effects. At 6:00 on August 25, 2015, when the typhoon was positioned off the west coast of Kyushu Island, there was a prevailing southwestward wind in the Tsushima Strait (Figure~\ref{fig:dist_T1515_w}a). This southwestward wind induced a southwestward flow (Figure~\ref{fig:dist_T1515_w}a'), causing a negative surge anomaly along the northern coast of Kyushu Island and a positive surge anomaly on the southern coast of the Korean Peninsula due to Ekman transport (Figure~\ref{fig:dist_T1515_w}a’). Therefore, based on wind effects alone, Hakata Bay would experience a -20~cm negative surge anomaly. 
However, at this time, there was a positive surge anomaly in Hakata Bay due to the inverted barometer effects. A positive surge anomaly due to the inverted barometer effects was canceled out by the negative anomaly due to Ekman transport, resulting in a net surge anomaly close to 0~cm (Figure~\ref{fig:cal_hakata_pw}d: p\&w).

As Typhoon T1515 passed directly over Hakata Bay at 8:30 on August 25, 2015, the boundary between the positive and negative surge anomalies shifted to the northern coast of Kyushu Island, and Hakata Bay experienced a wind-induced surge anomaly close to 0~cm (Figure~\ref{fig:dist_T1515_w}b''). The movement of the boundary is attributed to the strengthening of the southwestward wind (Figure~\ref{fig:dist_T1515_w}b), which enhances Ekman transport and causes the flow direction to shift to the west-southwest, pushing seawater toward Tsushima Island (Figure~\ref{fig:dist_T1515_w}b'). Considering the atmospheric pressure effects, the net surge anomaly in Hakata Bay sharply increased because of the cancelation of the wind-induced negative surge and pressure-induced positive surge, resulting in the first peak.
By 12:00 on August 25, 2015, as the typhoon moved northward, the situation reversed, and the boundary between the positive and negative surge anomalies returned to the northern coast of Kyushu Island (Figure~\ref{fig:dist_T1515_w}c''). The reason for the repositioning of the boundary could be the weakening of the wind, leading to a reduction in the flow and initiating oscillation, or the possibility that southeastward winds caused a reversal in the flow direction, returning it southeastward.

Conversely, focusing on the net surge anomaly in Hakata Bay during the second peak at 12:00 on August 25, 2015, the positive surge anomaly of 20~cm generated by the inverted barometer effect (Figure~\ref{fig:cal_hakata_pw}d: p) was offset by the wind-induced negative surge of -10~cm (Figure~\ref{fig:cal_hakata_pw}d: w), resulting in a second peak of 10~cm (Figure~\ref{fig:cal_hakata_pw}d: p\&w). However, the simulated value for the second peak was underestimated by 20~cm compared with the observed value (Figure~\ref{fig:compare_obc_cal}). This discrepancy could be attributed to the coarse computational grid in Hakata Bay and the omission of breakwaters within the bay, leading to an inaccurate calculation, and thus contributing to the second peak. Moreover, the influence of the wave setup, which was not considered in this simulation, may have also played a role. Therefore, addressing these limitations by refining the computational grid within Hakata Bay, accurately representing the bay's topography, and considering the impact of waves, could lead to a more precise numerical simulation of Hakata Bay. This would, in turn, enable a detailed understanding of storm surge development mechanisms for typhoons passing directly over the region, which represents a future research challenge.

Furthermore, we could not obtain high-quality meteorological data with high spatial resolution for the period before 1999. Although typhoons such as Typhoon Mireille and Seibolt Typhoon, classified as direct pass types, caused significant storm surge disasters in Hakata Bay, their storm surge development mechanisms were not investigated in this study. Elucidating the storm surge mechanisms of these typhoons will provide valuable insights for coastal disaster prevention in Hakata Bay. The Meteorological Agency began providing third-phase long-term reanalysis data (JRA-3Q) from September 1947 to August 2023, offering a high-quality and homogeneous dataset for an extended period. Therefore, future research should utilize such high-quality reanalysis data to conduct studies that consider the topographical conditions and wave effects in Hakata Bay.

\section{Conclusion}\label{Conclusion}
This study aims to elucidate the unique storm surge development mechanism on the northern coast of Kyushu, where limited investigations have been conducted to assess storm surges. Hakata Bay is located centrally on the northern coast of Kyushu Island, and the analysis involves the examination of observational data and numerical simulations along the Tsushima Strait. The following sections summarize the insights gained from this study for each typhoon tracking pattern.

\begin{enumerate}
    \item Northeastward-moving types
        \begin{itemize}
            \item Among the northeast-moving typhoons that pass near the south coast of the Korean Peninsula, the following trend was observed in Hakata Bay's storm surge. The bay experiences the first peak of storm surge during the typhoon's approach, followed by a rapid decrease to a minimum peak. After the typhoon moves away, the storm surge increases again, reaching a second peak. Subsequently, the storm surge slowly decreases with oscillations (T0314, T0415, T1618, T1825, T2211).
            \item For northeast-moving typhoons passing near the northern coast of Kyushu Island, Hakata Bay's storm surge tends to have only one peak (T1324, T1807, T1917).
            \item Spectral analysis of observational data revealed oscillations with periods of 2 hours, 7 hours, and 10 hours in Hakata Bay during typhoon events. The 2-hour oscillation is the harbor oscillation in Hakata bay, starting before the typhoon's approach and continuing afterward. The 7-hour oscillation corresponds to the rapid decrease and increase in storm surge after the first peak, and the 10-hour oscillation aligns with oscillations after the second peak, lasting for over 2 days.
            \item The 10-hour oscillation is identified as the Tsushima Strait's natural oscillation. It is a one-dimensional free oscillation with the boundary between the rapidly-deepening Sea of Japan and the East China Sea acting as nodes, and the north coast of Kyushu Island and the south coast of the Korean Peninsula as antinodes.
            \item When a typhoon approaches the Tsushima Strait, an inverted barometer effect occurs due to a pressure drop. The first peak is attributed to this. Subsequently, as the typhoon moves northeast, the rising pressure releases the restrained sea, leading to the generation of the 10-hour oscillation, causing the minimum and second peaks.
            \item The first peak due to wind is caused by the stagnation of seawater in the Tsushima Strait. Afterward, as the flow causing the stagnation disappears, positional energy is released, initiating oscillations and resulting in the minimum peak. The second peak in storm surge is attributed to Ekman transport and the oscillation in the Tsushima Strait.
            \item When wind rotating counterclockwise is applied, Ekman transport has a smaller peak compared to steady wind. Conversely, however, with wind rotating clockwise, the flow is bent by the Coriolis force, causing the wind direction to change more quickly than with steady wind. This leads to an earlier and larger peak.
            \item The peak in storm surge caused by typhoons passing near the northern coast of Kyushu or Kyushu Island is due to the circulation of counterclockwise flow formed around Tsushima Island, pushing seawater onto the northern coast of Kyushu Island and the south coast of the Korean Peninsula.
        \end{itemize}
    \item Northward-moving types
        \begin{itemize}
            \item Until the second peak, the development mechanism is similar to northeast-moving typhoons passing near the south coast of the Korean Peninsula.
            \item After the second peak, because northeastward winds continue to blow over the Tsushima Strait, Ekman transport maintains a high storm surge for an extended period on the northern coast of Kyushu Island.
        \end{itemize}
    \item Directly passing overhead types
        \begin{itemize}
            \item Storm surges associated with directly passing typhoons exhibit the characteristic of having one or two peaks, occurring in a short period around the time of the typhoon's closest approach.
            \item As the typhoon approaches Hakata Bay, the southwestward flow induced by the wind leads to negative surges on the northern coast of Kyushu Island. Moreover, as the typhoon moves even closer to Hakata Bay, seawater is constrained by Tsushima Island, leading to a reduction in the magnitude of the negative surge on the northern coast of Kyushu Island. Additionally, at the time of the closest approach, the positive surge by the inverted barometer effect became greater than the negative surge by wind, leading to a rapid peak in surges in Hakata Bay.
            \item The second peak is attributed to harbor oscillations generated by the typhoon passing through Hakata Bay.
            \item It should be noted that the simulation results tend to underestimate the magnitude of the second peak, and a more accurate assessment of the storm surge mechanism might require the consideration of coastal barriers and wave effects within the bay.
        \end{itemize}
\end{enumerate}

\section*{Acknowledgement}
  We would like to thank Editage (www.editage.jp) for English language editing.

\section*{Declaration of interests}
The authors declare that they have no known competing financial interests or personal relationships that could have appeared to influence the work reported in this paper.



\bibliography{cas-refs}

\end{document}